\documentclass[preprint,journal ]{vgtc}       

\ifpdf
  \pdfoutput=1\relax                   
  \usepackage{graphicx}                
  \DeclareGraphicsExtensions{.pdf,.png,.jpg,.jpeg} 
\else
  \usepackage{graphicx}                
  \DeclareGraphicsExtensions{.eps}     
\fi%

\graphicspath{{figures/}{pictures/}{images/}{./}} 

\usepackage{microtype}                 
\PassOptionsToPackage{warn}{textcomp}  
\usepackage{textcomp}                  
\usepackage{mathptmx}                  
\usepackage{times}                     
\usepackage{cite}                      
\usepackage{tabu}                      
\usepackage{booktabs}                  
\usepackage{color}
\usepackage[linesnumbered, ruled]{algorithm2e}
\usepackage{amsmath}
\usepackage{wrapfig}

\usepackage{graphicx}
\usepackage[labelfont=sf,textfont=sf]{subcaption}
\usepackage[pass]{geometry}
\usepackage{multirow}
\usepackage{makecell}
\usepackage{hyperref}
\usepackage{ccicons}

\usepackage{pgfplots}
\pgfplotsset{compat=1.16}
\usepackage[export]{adjustbox}

\usepackage{soul}

\onlineid{1477}

\vgtccategory{Research}
\vgtcpapertype{algorithm/technique}

\newcommand{\SystemName}{spEuler}
\title{\textsc{\SystemName}: Semantics-preserving Euler Diagrams}

\author{Rebecca Kehlbeck, Jochen Görtler, Yunhai Wang, and Oliver Deussen}
\authorfooter{
\item R. Kehlbeck, J. Görtler, O. Deussen are with the University of Konstanz, Germany. Email:  firstname.lastname@uni-konstanz.de
\item Yunhai Wang is with Shandong University, China. Email: cloudseawang@gmail.com
\item Oliver Deussen and Yunhai Wang are joint corresponding authors
}

\shortauthortitle{Firstauthor \MakeLowercase{\textit{et al.}}: Paper Title}

\abstract{Creating comprehensible visualizations of highly overlapping set-typed data is a challenging task due to its complexity. 
To facilitate insights into set connectivity and to leverage semantic relations between intersections, we propose a fast two-step layout technique for Euler diagrams that are both well-matched and well-formed. 
Our method conforms to established form guidelines for Euler diagrams regarding semantics, aesthetics, and readability. 
First, we establish an initial ordering of the data, which we then use to incrementally create a planar, connected, and monotone dual graph representation. 
In the next step, the graph is transformed into a circular layout that maintains the semantics and yields simple Euler diagrams with smooth curves. 
When the data cannot be represented by simple diagrams, our algorithm always falls back to a solution that is not well-formed but still well-matched, whereas previous methods often fail to produce expected results. 
We show the usefulness of our method for visualizing set-typed data using examples from text analysis and infographics. 
Furthermore, we discuss the characteristics of our approach and evaluate our method against state-of-the-art methods.%
} 

\keywords{Euler diagrams, Venn diagrams, set visualization, layout algorithm}

\teaser{
	\centering
	\begin{subfigure}[b]{0.305\linewidth} %
        \includegraphics[width =\textwidth]{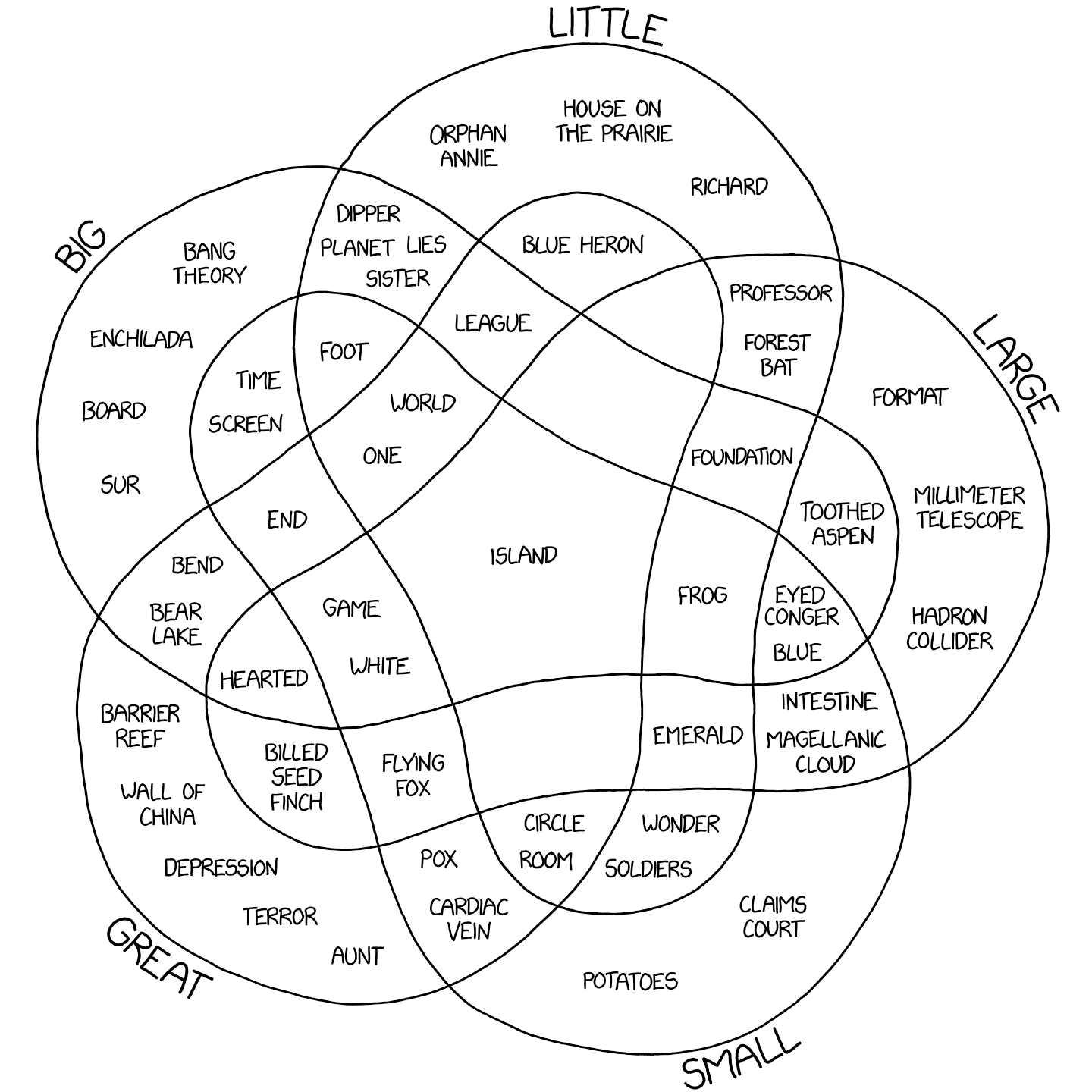}
        \caption{Original (\texttt{https://xkcd.com/2122/})}
        \label{fig:teaser:original}
    \end{subfigure}
    \begin{subfigure}[b]{0.3\linewidth}
        \includegraphics[width =\textwidth]{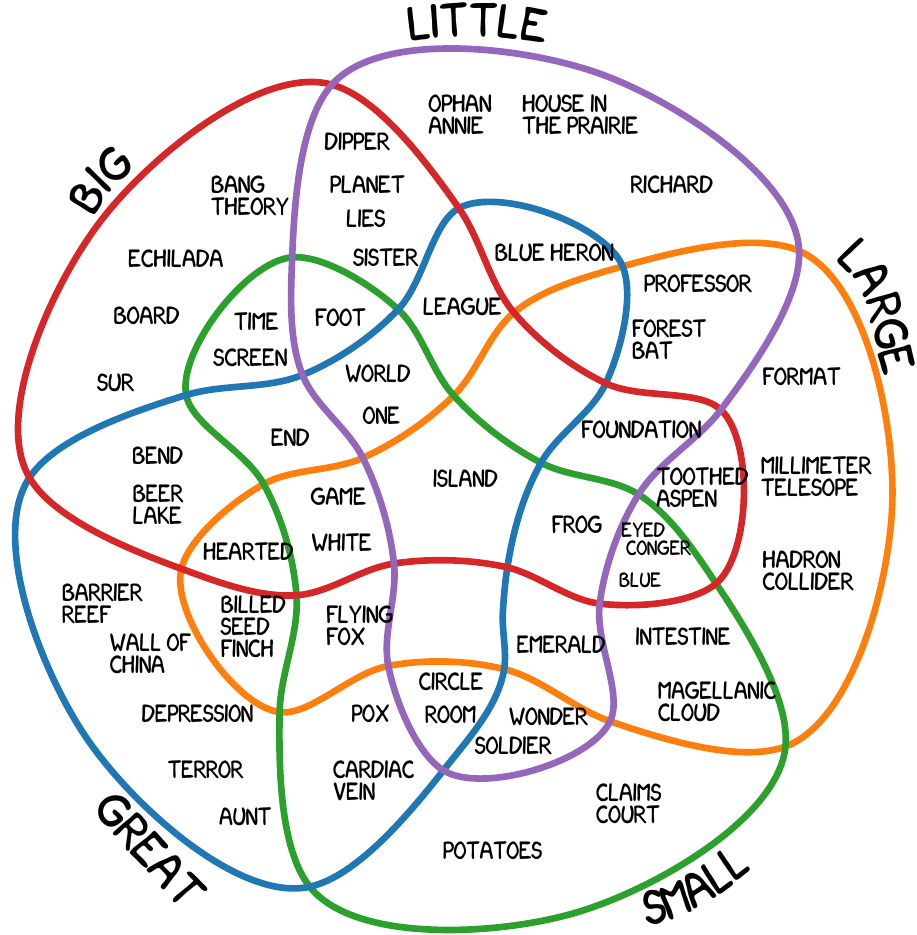}
        \caption{Reconstructed using our method}
         \label{fig:teaser:our-venn}
    \end{subfigure}
    \begin{subfigure}[b]{0.3\linewidth}
        \includegraphics[width =\textwidth]{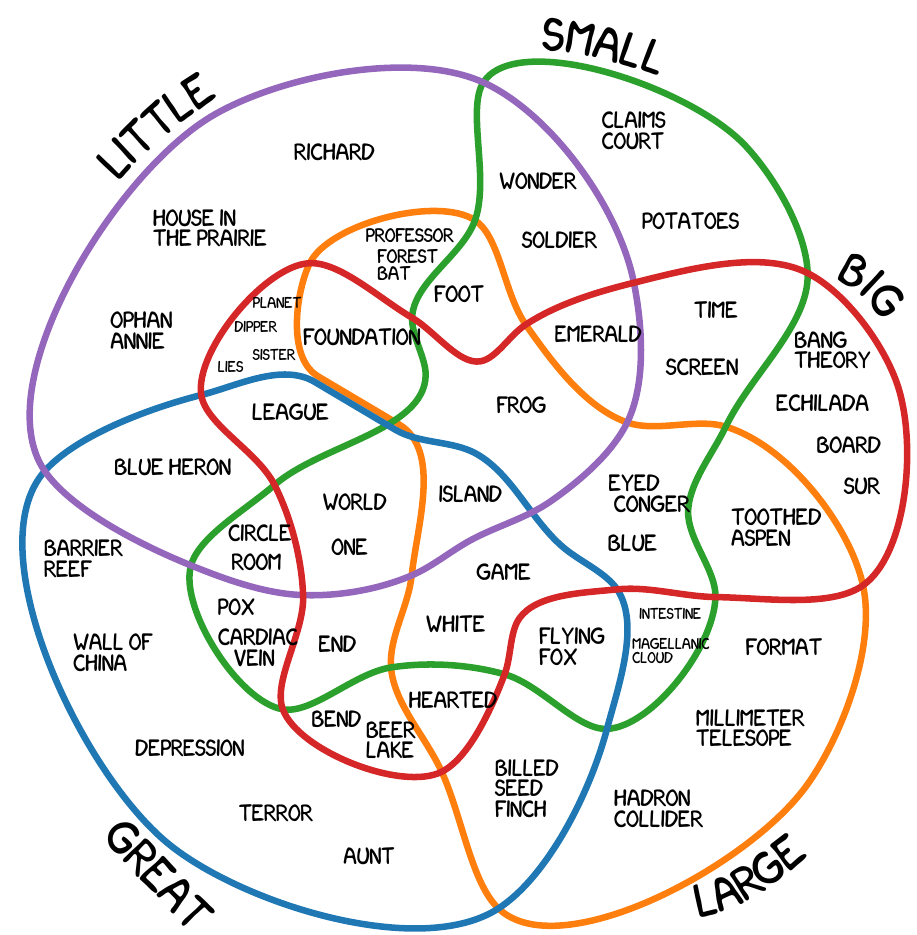}
        \caption{Well-matched result using our method}
         \label{fig:teaser:our-euler}
    \end{subfigure}
    \caption{Venn diagrams are often used to highlight complex interactions of sets.
    This example from \texttt{xkcd.com} shows which adjectives can be used in combination (a).
    Using our method, we can recreate this manually created Venn diagram (b).
    Here, the diagram contains empty intersections.
    In these cases, Euler diagrams (c) provide a more faithful representation of the data.
    \st{With our algorithm, we can create both Venn and Euler diagrams that satisfy established form guidelines.}
    }\label{fig:teaser}
}

\newif\ifshowcomments
\showcommentsfalse

\definecolor{blue(ncs)}{rgb}{0.0, 0.53, 0.74}

\ifshowcomments
\newcommand{\todo}[1]{{\color{red}{[TODO: #1]}}}
\newcommand{\TODO}[1]{{\color{red}{[TODO: #1]}}}
\newcommand\od[1]{{\color{cyan}#1}}

\newcommand{\yh}[1]{{\color[rgb]{0.7,0.7,0}{[YH: #1]}}}
\newcommand{\revised}[1]{{\color{blue(ncs)} [#1]}}
\newcommand{\rk}[1]{{\color{blue} [RK: #1]}}
\newcommand{\jg}[1]{{\color[rgb]{1,0.5,0} [JG: #1]}}

\newcommand{\ma}[1]{{\color{green} [MA: #1]}}
\newcommand{\pp}[1]{{\color{violet} [PP: #1]}}
\else
\renewcommand{\st}[1]{}
\newcommand{\todo}[1]{}
\newcommand{\TODO}[1]{}
\newcommand{\revised}[1]{#1}
\newcommand{\yh}[1]{}
\newcommand{\rc}[1]{}
\newcommand{\od}[1]{}
\newcommand{\rk}[1]{}
\newcommand{\jg}[1]{}
\newcommand{\ma}[1]{}
\newcommand{\pp}[1]{}
\fi

\definecolor{blueNode}{RGB}{60,117,175}
\definecolor{redNode}{RGB}{196,59,54}
\definecolor{orangeNode}{RGB}{240,144,69}

\newcommand{\nodeIcon}[2]{%
    \begin{tikzpicture}[scale=0.5, baseline=-1mm, thick]
        \node[scale=0.6, minimum size=0.55cm, inner sep=0pt, fill={#1}, circle] {\textsf{\textbf{\textcolor{white}{#2}}}};
    \end{tikzpicture}%
}

\newcommand{\PropWellMatched}{\textbf{P1}}
\newcommand{\PropWellFormed}{\textbf{P2}}
\newcommand{\PropConvex}{\textbf{P3}}
\newcommand{\PropSmoothCurves}{\textbf{P4}}
\newcommand{\PropSymmCurves}{\textbf{P5}}
\newcommand{\PropShapeDisc}{\textbf{P6}}
\newcommand{\PropZoneArea}{\textbf{P7}}

\newcommand{\Set}[1]{#1}

\newcommand{\eg}{e.g.}

\newcommand{\FlagIcon}[1]{%
    \begin{tikzpicture}[scale=0.4, baseline=-0.9mm]
        \node[scale=0.17] {\includegraphics[trim=1mm 1mm 0mm 0mm, clip]{#1}};
    \end{tikzpicture}%
}

\newcommand{\FlagColombia}{\FlagIcon{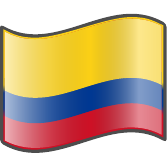}}
\newcommand{\FlagCostaRica}{\FlagIcon{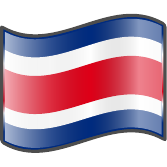}}

\begin{document}
 \maketitle

\abstract{Creating comprehensible visualizations of highly overlapping set-typed data is a challenging task due to its complexity. 
To facilitate insights into set connectivity and to leverage semantic relations between intersections, we propose a fast two-step layout technique for Euler diagrams that are both well-matched and well-formed. 
Our method conforms to established form guidelines for Euler diagrams regarding semantics, aesthetics, and readability. 
First, we establish an initial ordering of the data, which we then use to incrementally create a planar, connected, and monotone dual graph representation. 
In the next step, the graph is transformed into a circular layout that maintains the semantics and yields simple Euler diagrams with smooth curves. 
When the data cannot be represented by simple diagrams, our algorithm always falls back to a solution that is not well-formed but still well-matched, whereas previous methods often fail to produce expected results. 
We show the usefulness of our method for visualizing set-typed data using examples from text analysis and infographics. 
Furthermore, we discuss the characteristics of our approach and evaluate our method against state-of-the-art methods.} 

\keywords{Multi-class visualization, layout, Venn diagrams, Euler diagrams, chain decomposition}

\firstsection{Introduction}

\maketitle

\section{Introduction}

Set-typed data is ubiquitous across many different research areas, such as multi-label classification~\cite{wei2015hcp} in machine learning, RNA and DNA sequencing~\cite{ramirez2018high, hentze2018brave, d2012banana} in computational biology, and topic modeling~\cite{blei2003latent} in natural language processing.
There are two prominent methods to visualize set relations.
Venn diagrams~\cite{venn1880diagrammatic} show all possible relations between sets.
In contrast, Euler diagrams \cite{leonhard1768lettres} only depict non-empty relations\st{ and therefore preserve the semantics of the data faithfully}.
Many special-purpose visualizations have been developed for set-specific tasks~\cite{alsallakh2014visual}.
Still, traditional Venn and Euler diagrams remain an essential tool for showing set intersections because they are easy to read, familiar to most users, and can incorporate data points directly.
As such, they are often part of larger systems, such as \textit{UpSet}~\cite{LexGSVP14}.

Due to their combinatorial nature, the construction of Venn diagrams is straightforward.
However, automatically creating Euler diagrams of high quality remains a challenging task, in particular for highly intersecting datasets.
An Euler diagram should only include relations that are present in the data and avoid introducing superfluous areas.
Further, the diagram should be monotone~\cite{Cao2010}.\st{, which means that there are only pairwise intersections of curves and, if possible, no curves are concurrent to another.}
We call Euler diagrams that adhere to these properties \textit{semantics-preserving}, following the definition of semantics in the domain of linguistics. 
Accordingly, representing the data faithfully and preserving neighbourhood relations are a part of semantics, as how a set intersection is read depends on its neighbours.
An example result of our method and the impact of the above-mentioned properties is shown in \autoref{fig:teaser:our-euler}.
The Euler diagram on the right has lost the symmetry of the Venn diagram (\autoref{fig:teaser:our-venn}) but represents the data faithfully\st{ by removing empty intersections}. 
\st{As each of the diagrams may be preferred by users, depending on the task and what information they want to emphasize, our method is able to provide both solutions.}

First, we introduce and formalize the properties of Euler diagrams.
Next, we propose a two-step algorithm for constructing such diagrams efficiently.
The first step computes the \textit{Euler dual}, a graph representation of the \st{final} diagram.
\st{From the dual, t}The second step \st{of the algorithm then} creates \st{curves}\revised{the \textit{Euler diagram}}, whose curves follow guidelines~\cite{Blake2016} for creating intuitive \st{and readable} Euler diagrams.
We show the usefulness and characteristics of our algorithm on three examples from different domains and \st{evaluate}\revised{compare} our method to previous work.
In summary, the main contributions of this paper are:

\begin{itemize}
    \setlength\itemsep{0.25em}
    \item \textsc{\SystemName}, a \textbf{novel method} for constructing semantics-preserving Euler diagrams that yield fast and reliable results.
    \item Extensive \textbf{analysis of existing construction methods} and how they relate to properties of the Euler diagrams.
    \item \textbf{Three examples} from different domains that show the characteristics and potential of our approach.
    \item An \textbf{extensive evaluation} based on established guidelines of Euler diagrams and direct comparison to state-of-the-art methods.
\end{itemize}

\section{Characteristics of Euler Diagrams}\label{sec:properties}
Before we go into the previous work that is related to our method, we want to introduce important properties and concepts of Euler diagrams that will help to understand the subsequent sections.
Formally, an Euler diagram is a set of smooth, closed Jordan curves that represent the different sets~\cite{chow2007generating}.
Together, these curves comprise various areas in the drawing that represent the intersections of the sets. 
All set relations that exist in the data can be described by the \textit{abstract description}---a list of the existing intersections.
Euler diagrams can exhibit several different properties that directly influence their appearance and effectiveness in visualizing information.
The two most important properties are well-formedness and well-matchedness, as defined by Chow~\cite{chow2007generating}.

\begin{figure}
    \centering
    \begin{subfigure}[b]{0.45\linewidth}
        \includegraphics[width=0.9\linewidth]{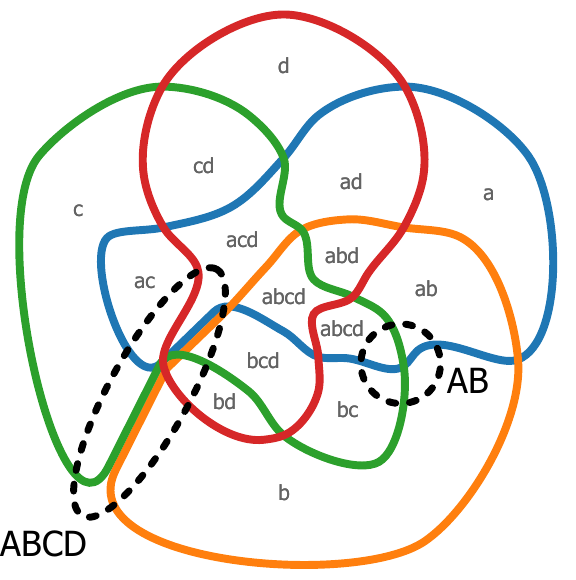} 
        \caption{well-matched}
        \label{fig:properties:well-matched}
    \end{subfigure}
    \begin{subfigure}[b]{0.45\linewidth}
        \includegraphics[width=0.9\linewidth]{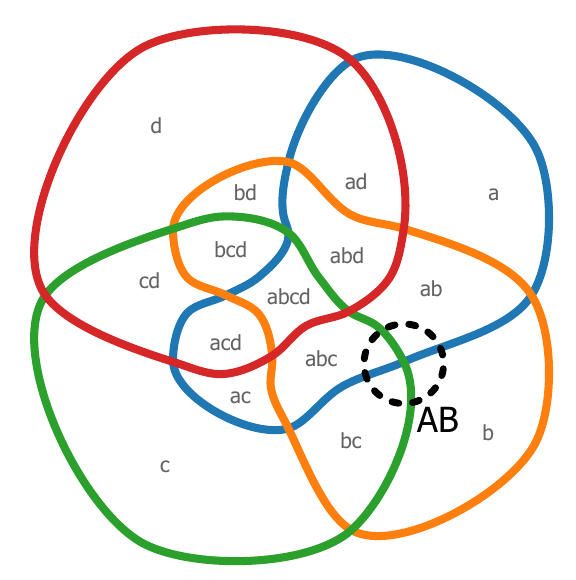}
        \caption{well-formed and well-matched}
        \label{fig:properties:well-formed}
    \end{subfigure}
    \begin{subfigure}[b]{0.45\linewidth}
        \includegraphics[width=0.98\linewidth]{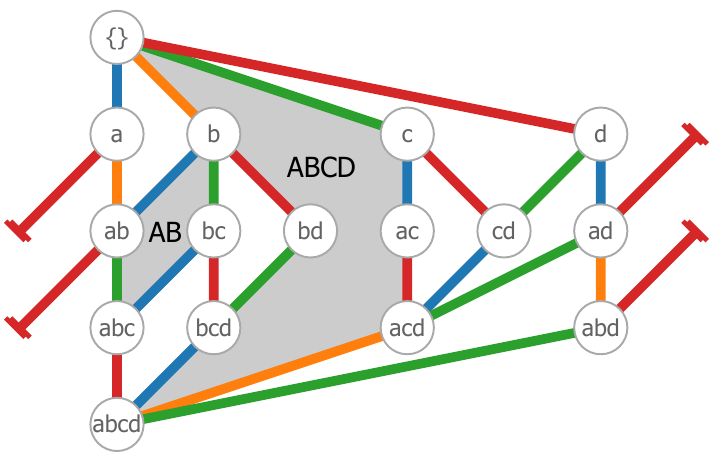}
        \caption{Euler dual of (a)}
        \label{fig:properties:duals:well-matched}
    \end{subfigure}
    \begin{subfigure}[b]{0.45\linewidth}
        \includegraphics[width=0.98\linewidth]{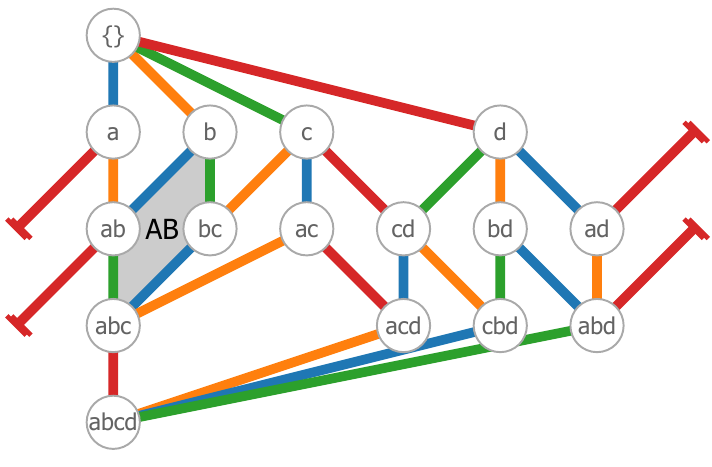}
        \caption{Euler dual of (b)}
        \label{fig:properties:duals:well-formed}
    \end{subfigure}
    \caption{
        (a) A well-matched diagram and (b) an additionally well-formed diagram.
        Well-matched diagrams may exhibit concurrent curves and points where more than two curves intersect, \eg, the intersection of curves $\Set{ABCD}$.
        On the other hand, well-formed diagrams do not have these problems and only have pairwise intersections, \eg, $\Set{AB}$.
        (c) and (d) show the ranked-based duals for (a) and (b).
        The concurrency surfaces as face $\Set{ABCD}$ in (c).
        The well-formed diagram instead only contains faces with 4 surrounding links. 
        We will explain the impact of this in \autoref{sec:euler-dual}.
    }\label{fig:properties}
\end{figure}

\paragraph{Properties}
An Euler diagram is \textbf{well-formed}, if it is \textit{simple} (i.e. at most, two curves meet at any given point and there is no concurrency), and exactly a \textit{single curve} represents each set.
In a \textbf{well-matched} Euler diagram, all intersections are correctly represented, thereby retaining the semantics from the original data: each intersection is represented only once, and the diagram does not contain areas of intersections that are not part of the abstract description.
Alsallakh et al.~\cite{alsallakh2014visual} discuss different properties of algorithms for Euler diagrams and their connection to well-formedness.
However, there is no such discussion for the well-matchedness and the interplay between both properties, which plays a big role in the effectiveness of the diagram~\cite{Gurr99}.
The two properties are visualized in \autoref{fig:properties}, which shows a Venn diagram with 4 curves and their 16 intersections.
We use uppercase letters to refer to a curve or all nodes that participate in a set, and lowercase letters to refer to specific intersections, which are faces (also called \textit{zones}) in the diagram.
We will revisit this simple example throughout the next sections to help showcase our method. 
\autoref{fig:properties} shows the visual differences of adhering to only one or both of these two properties for the same data. Each zone is marked with its respective intersection.
As can be observed in \autoref{fig:properties:well-matched}, all four curves intersect on the lower-left corner, resulting in concurrent lines. 
By creating a well-matched and well-formed diagram, this can be avoided (\autoref{fig:properties:well-formed}).
It is important to note that many abstract descriptions exist, for which both properties cannot be satisfied at the same time, requiring a trade-off.
However, as analyzed by Chow~\cite{chow2007generating}, it is currently not possible to infer for a given abstract description if it is possible to maintain both properties.
If a trade-off has to be made, we adhere to the guidance of the work by Chapman et al.~\cite{ChapmanSRMB14}, 
which concludes that users prefer well-matched diagrams over well-formed ones.
As a result, in these cases, our algorithm always produces well-matched diagrams while minimizing the violations of well-formedness.

\paragraph{Euler Dual}
A key concept that frequently shows up in construction algorithms is modeling the Euler diagram as a graph.
Instead of thinking about the Euler diagram as a set of curves, it can be modeled directly as an edge-labeled graph, called the \textit{Euler graph}.
In this representation, each intersection of the curves is represented by a node, and each curve segment is represented by a link, labeled with the respective curve of the underlying original Euler diagram.
Instead of creating the Euler diagram directly from the data using curves, it is also possible to indirectly create it by constructing the \textbf{Euler dual} of the Euler graph.
Each node in the Euler dual represents a face of the Euler graph, and neighboring zones are represented by linked nodes in the Euler dual. However, in theory, all nodes that differ by one set could be linked in the dual---a graph that contains all possible links is therefore called the \textit{super dual}.
The \textit{rank} of a node in the Euler dual equals the number of sets participating in that intersection.
We can find an ordered representation of the Euler dual by grouping all nodes of the dual that have the same rank. The resulting graph is the \textit{rank-based Euler dual}. \autoref{fig:properties:duals:well-matched} and \autoref{fig:properties:duals:well-formed} show the respective rank-based duals of \autoref{fig:properties:well-matched} and \autoref{fig:properties:well-formed}---the non-pairwise intersection of \autoref{fig:properties:well-matched} is equal to the face $\Set{ABCD}$ in \autoref{fig:properties:duals:well-matched}. In comparison, all the faces of \autoref{fig:properties:duals:well-formed} are quads---we will explain what this means for the diagram in \autoref{sec:euler-dual}.

\begin{table*}[htb]
    \small
    \centering
    \caption{Details of different construction methods for any amount of curves and their properties.}\label{table:Properties}
    \begin{tabular}{lcccccccc} 
        \toprule
        Method & Construction & \makecell{Any \\ relation} & \makecell{\bf Well-\\\bf matched} & \makecell{\bf Well-\\\bf formed} & \makecell{Monotonicity} & simple & \makecell{Duplicate \\ curves} & \makecell{Non-pairwise \\ intersections} \\
        \toprule
        SCD~\cite{ruskey2006search} / nVenn~\cite{Perez-SilvaAQ18} & Euler dual & yes & no & no             & yes & no & no & yes\\
        Stapleton\revised{~\cite{StapletonFRH12}}/ Rodgers~\cite{RodgersSAMBT16} & direct                       & yes & no & no   & yes & yes & yes & no\\
        \midrule
        Venn~\cite{venn1880diagrammatic} & direct                               & yes & no & \textbf{yes}   & yes & yes & no & no\\ 
        Edwards~\cite{edwards1989venn} & direct                                 & yes & no & \textbf{yes}   & yes & yes & no & no \\
        vennEuler~\cite{Wilkinson2012} & direct                                 & no & no & \textbf{yes}    & yes & yes & no & no\\
        eulerr~\cite{Larsson2020} & direct                                      & no & no & \textbf{yes}    & yes & yes & no & no \\
        \midrule
        Chow-Ruskey~\cite{ChowR05} & Euler dual                       & yes & \textbf{yes} & no   & yes & no & no & yes \\
        Simonetto~\cite{SimonettoA09} & Intersection graph                      & yes & \textbf{yes} & no   & yes & yes & yes & no\\
        MetroSets~\cite{Jacobsen2021} & hypergraph                              & yes & \textbf{yes} & no   & no & - & - & yes\\
        \midrule
        Flower~\cite{FlowerH02,FlowerFH08}\textsuperscript{\textasteriskcentered} &  Euler dual & no & \textbf{yes}  & \textbf{yes} & yes & yes & yes & no\\
        \textbf{Our method} &  Euler dual & no & \textbf{yes}  & \textbf{yes}                               & yes & yes & yes & no\\
        \bottomrule 
    \end{tabular}
    \parbox{0.85\textwidth}{\footnotesize
    \small
        \textsuperscript{\textasteriskcentered}Note: The authors only provide a rough sketch of their method.
    } 
\end{table*}

\section{Related Work}\label{sec:related}

Many set visualization approaches have been proposed in the past. Good starting points are the survey of Venn diagrams by Ruskey and Weston~\cite{ruskey1997survey}, or Rodgers~\cite{Rodgers14}, who focuses on Euler diagrams. Alsallakh et al.~\cite{alsallakh2014visual} offer a comprehensive survey of set visualizations and group the techniques based on their best-suited tasks: Element tasks, set relation tasks, and element attributes tasks.

\subsection{General Set Visualization}

Alternative approaches to visualize set-typed data are matrix and aggregation-based techniques, such as UpSet~\cite{LexGSVP14} or RadialSets~\cite{AlsallakhAMH13}. 
These are usually very well suited for element and element attribute tasks. However, they can be verbose to show all set relations at once when the data is complex. 

For spatial data, such as maps, there are also techniques that focus on highlighting the connections between sets, such as BubbleSets~\cite{collins2009Bubblesets} or KelpFusion~\cite{MeulemansRSAD13}. 
Most methods are not able to directly encode information of the original data points in a unified visualization. 
For this task, Venn and Euler diagrams are especially well suited and therefore have been combined with glyphs~\cite{MicallefDF12}, and graphs~\cite{MuttonRF04, SathiyanarayananSBH14}.
Finally, Jacobsen et al.~\cite{Jacobsen2021} propose using the metro map metaphor to visualize set relations in their MetroSets technique.
The visualization can show individual data points for each set relation, and the layout can be fine-tuned according to different optimization strategies.

\subsection{Constructing Venn and Euler Diagrams}
Venn diagrams always show all possible set relations, with many different methods for their construction~\cite{venn1880diagrammatic, edwards1989venn, ruskey2006search, Rodgers14, Bannier2017}.
Euler diagrams are more flexible in this regard, but many construction algorithms are limited to specific abstract descriptions and might produce unexpected results~\cite{chow2007generating, RodgersZF08, FlowerFH08,micallef2014eulerforce, simonetto2016simple}.

Inductive methods construct diagrams by adding one curve at a time. 
Venn himself proposed an inductive method to create diagrams for any amount of curves.
Edwards later proposed an alternative inductive construction method that creates diagrams by projecting the curves onto a sphere~\cite{edwards1989venn}.
This method always creates diagrams that are well-formed and well-matched. However, for a larger number of sets, the result becomes hard to understand as the area of new zones becomes smaller and smaller.
Other methods focus on the creation of simple, convex Venn diagrams, \eg, Mamakani et al.\cite{MamakaniMR12}, which are aesthetically more pleasing.
Ruskey et al.~\cite{ruskey2006search} use a general Venn construction method to analyze methods that create symmetric Venn diagrams.
nVenn~\cite{Perez-SilvaAQ18}, a recently developed area-preserving Euler-like visualization technique, allows users to get a compact overview, even for larger set counts. 
They use a conventional Venn construction algorithm~\cite{ruskey2006search} as its initial layout and adapts it using a force-directed optimization. It heavily relies on the initial position\revised{ing and}\st{s during construction, as well as the} parameters of the force-directed strategy.

If the given dataset does not cover all possible set relations, Venn diagrams produce\st{, as mentioned above,} additional (unwanted) zones, and the diagram is not well-matched.
For diagrams that are not well-matched, there is a discrepancy between the semantically correct representation of the abstract description and the visualization.
Oftentimes, this problem is solved using shading to mark such additional faces~\cite{venn1880diagrammatic, StapletonFRH12}. 
In any case, this encodes unnecessary information that the reader has to process. A solution to this mismatch is well-matched Euler diagrams. 

By design, methods that create Euler diagrams are usually well-matched. 
Their drawback, however, is that they often cannot make any guarantee about the aesthetics of the diagrams, i.e., their well-formedness. 
Results might contain crossings, concurrent curves, and non-smooth shapes. 
To alleviate this problem, Stapleton et al.~\cite{StapletonFRH12} proposed an inductive method to create (semi) well-formed Euler diagrams using circles. 
Such diagrams weaken the constraints of the well-formedness and allow curve labels to be used multiple times. 
A current hindrance in the application of Euler diagrams is that most methods only produce expected results for certain datasets. 
Users do not know beforehand which method will produce well-formed or well-matched diagrams or if it will produce a valid result at all. 
\st{It is important to note that most e}Existing implementations often fail silently without producing any results or create unwanted zones without communicating this to the user.

It is challenging to create a well-formed and well-matched diagram for any abstract description because of the intricate interplay between the different properties.
Therefore, many construction methods that only optimize for one property often cannot make guarantees for the others.
This can be seen in \autoref{table:Properties}.
Usually, an Euler diagram is either \textit{directly} constructed via curves or \textit{indirectly} through an intermediate representation, which is then transformed into the Euler diagram.
Examples are constructions using the Euler dual, Euler graph, connectivity graph, closeness graph, or intersection graph.
Based on the surveys by Ruskey~\cite{ruskey1997survey}, Rodgers et al.~\cite{Rodgers14}, and Alsallakh et al.~\cite{alsallakh2014visual}, we created \autoref{table:Properties}, in which we compare different properties of Euler and Venn construction methods.

It should be noted that the properties of the final Euler diagram highly depend on the used construction steps as well as the properties of the intermediate representations.
As we can observe from the table, direct construction methods usually produce well-formed diagrams, as they directly model the curves.
This means the produced curves are usually constrained heavily, for example, by only using circles.
As a trade-off, they only produce Venn diagrams or introduce unwanted zones for higher set counts.
Alternatively, indirect methods only create the exact intersections needed and then transform the graph to the diagram but fail to create well-formed diagrams from them.
Some methods try to transform non-well-formed diagrams into more aesthetic ones, but doing this in hindsight is often not possible.
Examples can be found in~\cite{StapletonRHZ11, SimonettoA09, RodgersZF08, StapletonFRH12}.
There is only a single approach that allows for the creation of Euler diagrams of any amount of curves that are both well-matched and well-formed. 
Flower et al.~\cite{FlowerFH08} propose an initial sketch of a solution but do not propose a general implementation. 
They resort to heuristics to create solutions for less than 5 curves.
There are two main differences between our algorithm and the approach by Flower et al.~\cite{FlowerFH08}: They do not use the rank-based dual as an intermediate, and they cannot fall back to a sub-optimal solution when no well-formed and well-matched result exists.

\subsection{Evaluation of Euler Diagrams}
As mentioned previously, the properties of Euler diagrams can be generally divided into well-matched and well-formed diagrams.
However, there are many more properties that influence the semantics (e.g., monotonicity) and the aesthetics (e.g., shape, color, and symmetry). 
A general overview is given by Blake et al.~\cite{Blake2016}, which introduces different guidelines that good Euler diagrams should adhere to. 
\revised{They directly compare real-word examples with adapted diagrams, which follow their proposed guidelines. 
Comparing both, such diagrams improve user comprehension. 
However,
it is still unknown which of the guides might have a larger impact, and how they might influence each other.}
\st{Further, t}\revised{T}here are several studies that analyze the readability of well-matched vs. well-formed diagrams~\cite{ChapmanSRMB14, wallinger2021readability}. \st{and how diagrams can be combined with graphs~\cite{RodgersSAMBT16} to solve tasks concerning set visualizations.}
\revised{Chapman et al.~\cite{ChapmanSRMB14} compare various types of set diagrams
\st{ regarding task completion time and error rates}
and found that linear diagrams outperform all other methods, followed by unshaded Euler diagrams. 
They explain their results by the well-matchedness of those approaches, combined with well-formedness as a secondary influence. 
Rodgers et al.~\cite{RodgersSAMBT16} evaluate methods that combine the Euler diagram with a graph of the datapoints. As their results are not consistent with previous studies of the same methods, they suggest that this might be due to them using datapoint specific tasks, whereas the previous studies used intersection related tasks. 
They conclude that for graph specific tasks, 
the properties that we summarize as ``semantics preserving'' may 
explain why some methods perform better than others.} 
Wallinger et al.~\cite{wallinger2021readability} compare Euler diagrams with MetroSets and \revised{LineSets}\st{BubbleSets} for set-related tasks.

To conclude: it is still an open problem to design and implement an algorithm that produces well-matched and well-formed Euler diagrams for any amount of curves if the abstract description allows for it. 
Generating Euler diagrams with specific properties was also identified as an open problem by Alsallakh et al.~\cite{alsallakh2014visual}.
Depending on the existing relations in the data, some properties are impossible to guarantee.
We therefore propose a semantics-preserving construction method that generates Euler diagrams for any amount of curves.
It creates well-matched and well-formed diagrams if allowed for by the data.
If not, we retain the well-matchedness and relax as few individual properties as possible that infringe the well-formedness.
\begin{figure*}[ht!]
    \centering
    \begin{subfigure}[b]{0.16\textwidth}
        \includegraphics[width=\linewidth]{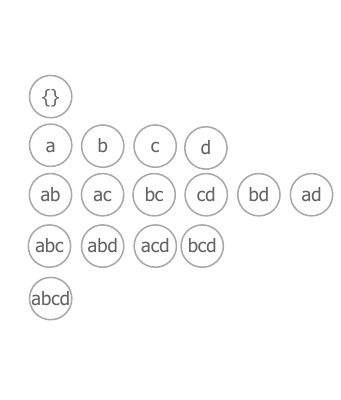}
        \caption{Abstract description}
        \label{fig:overview:abstract-description}
    \end{subfigure}
    \begin{subfigure}[b]{0.34\textwidth}
        \includegraphics[width=\linewidth]{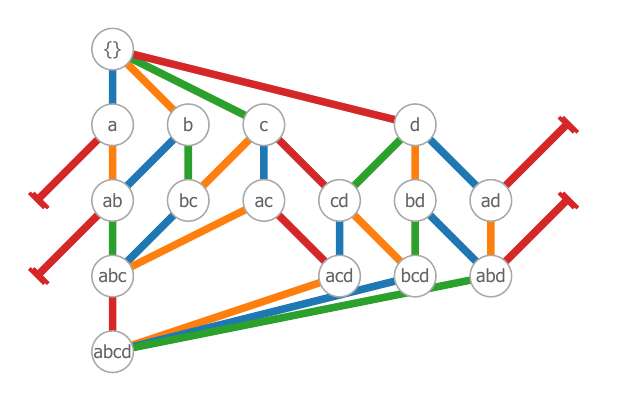}
        \caption{Rank-based Euler dual}
        \label{fig:overview:rank-based-dual}
    \end{subfigure}
    \begin{subfigure}[b]{0.2\textwidth}
        \includegraphics[width=\linewidth]{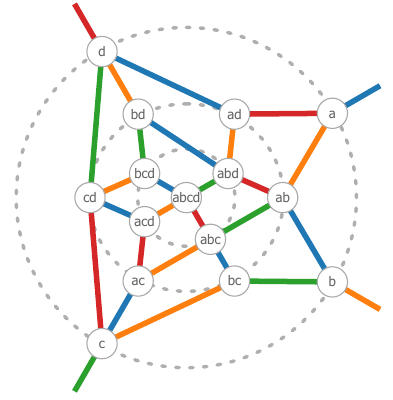}
        \caption{Circular layout}
        \label{fig:overview:circular-dual}
    \end{subfigure}
        \begin{subfigure}[b]{0.25\textwidth}
        \includegraphics[width=\linewidth]{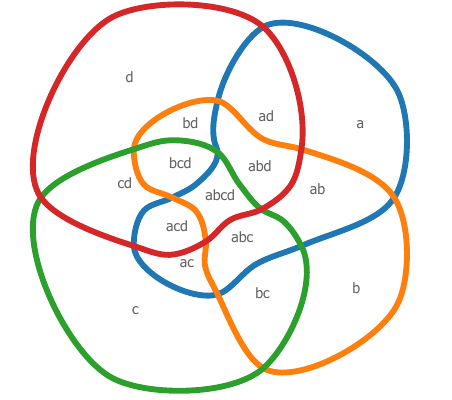}
        \caption{Final diagram}
        \label{fig:overview:euler-diagram}
    \end{subfigure}
    \caption{
    Overview of our method: After finding all set intersections that exist in the dataset (a), the rank-based Euler dual is created from the abstract description (b). The graph is then transformed to be circular, and nodes are arranged in a well-distributed manner across several rings (c). In (d) we create the final curve for each set using splines.}
    \label{fig:overview}
\end{figure*}
\section{Overview}\label{sec:overview}
Our method constructs Euler diagrams for a given list of sets and their intersections---the abstract description.
For example, the three sets $\{ \Set{A}, \Set{B}, \Set{C} \}$ can have relations
$\{ \emptyset,\Set{a},\Set{b},\Set{c},\Set{ab},\Set{bc},\Set{ac},\Set{abc}\}$, where we abbreviate the zone $\Set{A} \cap \Set{B} \cap \Set{C}$ with $\Set{abc}$.
However, in real-world data usually not all intersections are realized, for example, the intersection $\Set{ac}$ could be missing.
For some abstract descriptions, it is possible to find well-matched and well-formed diagrams. 
However, there are also many configurations where this is not possible---in these cases our algorithm yields well-matched diagrams, while minimizing the violations of the well-formedness property.
We provide further discussion on the influence of the abstract description on these properties \st{of the diagram} in \autoref{sec:discussion}.

Our algorithm consists of four main steps, see also \autoref{fig:overview}.
Starting from the \textbf{abstract description} (\autoref{fig:overview:abstract-description}), we first find the appropriate order in which we place each set.
Our algorithm then iteratively grows the graph based on this order while ensuring that new nodes conform to the well-formed property.
After finding the connected, planar \textbf{rank-based Euler dual} (\autoref{fig:overview:rank-based-dual}), our algorithm arranges the nodes in a \textbf{circular layout} (\autoref{fig:overview:circular-dual}), which we then use to draw the curves that correspond to each set.
Because of the properties of the dual, it is possible to generate a planar \textbf{Euler diagram} (\autoref{fig:overview:euler-diagram}) from this circular layout.
We use smooth curves to create compact and simple shapes.
In contrast to other techniques, we guarantee a semantic match between the data and the final diagram. In addition, by creating mostly simple set curves and diagrams, we support the readability of the diagram, avoiding unnecessary crossings and concurrency of curves.

To demonstrate the usefulness and evaluate the characteristics of our algorithm, we implemented a prototype in JavaScript and D3\footnote{D3: \texttt{\url{https://www.d3js.org}}}.
This implementation also allows stepping through the individual steps of our algorithm.
The prototype shows exemplary abstract descriptions that can be found in the paper, as well as different interactions that support set-related tasks such as visual identification of subsets and hovering.
The implementation of our prototype, together with the example datasets, can be found online\footnote{ \url{https://github.com/RKehlbeck/spEuler}}.
\begin{figure*}[tbh]
     \includegraphics[width =\linewidth]{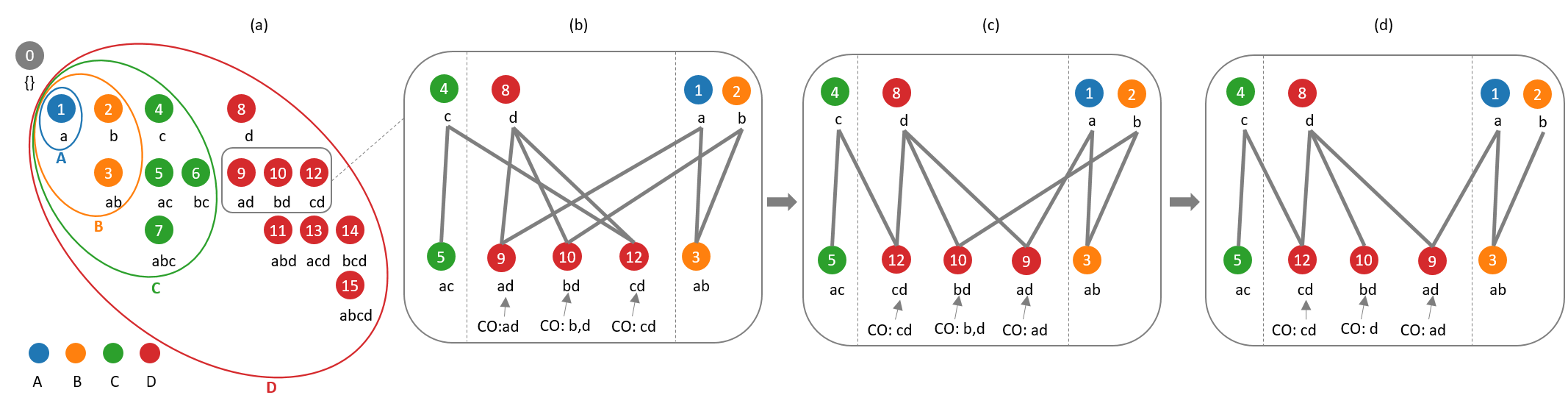}
    \caption{
        Creating the Euler dual:
        (a) shows the initial order of nodes and their respective groups.
        (b) Simply inserting the nodes with this initial order results in a dual that is non-planar.
        (c) We first remove crossings by changing the insertion order.
        (d) We finalize the graph by choosing the consecutive ones sequence which does not destroy monotone faces.
        The final result is a planar graph.
    }\label{fig:insertion}
\end{figure*}

\begin{algorithm}[t]
    \DontPrintSemicolon{}%
    \SetKwFunction{createDual}{createDual}%
    \SetKwProg{fn}{function}{}{}%
    \fn{\createDual{$ nodes[] $}}{
        $ G \gets $ group $nodes$ by extended set and rank\; \label{algo:grouping:1}
        $ G \gets $ sort $G$ by $rank(G_{i_{0}})$ and $len(G_{i})$ \; \label{algo:grouping:2}
        \ForAll{$S$ in $G$}{
            $R \gets$ group $S$ by rank \; \label{algo:CO:1}
            \ForAll{$r$ of $R$}{
                ${cos} \gets$ Calculate COs for $r$ \; \label{algo:CO:2}
                $r \gets $ sort $r$ by len($cos$) and $dist\_twin$ \; \label{algo:CO:3}
                \ForAll{$n$ of $r$}{
                    $list \gets []$ \;
                    \ForAll{co of $cos_{n}$}{
                        \ForAll{possible position p of co}{
                            $pp \gets \{len(co), \#mono, \#cross\}$ \; \label{algo:insert:1}
                            list.push(pp) \;
                        }
                    }
                    sort $list$ to maximize monotone faces\;
                    insert\_node($list$[0])\; \label{algo:insert:2}
                    remove\_crossings() \; \label{algo:insert:3}
                }
            }
        }
    }
    \caption{Dual construction algorithm}
    \label{alg:sort-nodes}
\end{algorithm}

\section{Rank-based Euler dual}\label{sec:euler-dual}

We already introduced the rank of an intersection in \autoref{sec:properties}, which is the number of its involved or participating sets.
In the context of the Euler dual, we will call these intersections \textit{nodes}.
For example, the node $\Set{a}$ has rank $1$, while node $\Set{ab}$ has rank $2$.
In the rank-based Euler dual, there can only be a link from a node with rank $r$ to a node with rank $r+1$.
This means that with each link, an additional set gets added to the intersection.
In the example above, this is the set $\Set{b}$, which we call the \textit{color} of a link.
By definition, a link always has a single distinct color.
Accordingly, we color the links in the figures containing Euler duals throughout this paper.

Our goal is to draw Euler diagrams with minimal violations of the well-formedness property.
As we create our diagram using the dual, we need to find the equivalent property in the dual that guarantees a well-formed result---the faces of the dual.
A face in the Euler dual is an area that is enclosed by links and nodes.
\textit{Monotone faces} are enclosed by exactly four links that have two alternating colors.
It is also important to note that by this definition, monotone faces always span exactly three ranks in the Euler dual.
Monotone faces are essential for well-formedness, as they limit how many curves can intersect at a given face---this directly corresponds to how simple the final Euler diagram will be.
\autoref{fig:properties} shows an example of a monotone and a non-monotone face.
Our goal is therefore to \textit{remove} possible links and \textit{reorder} the nodes across all ranks until we are left with a \textit{connected}, \textit{crossing-free} and \textit{monotone} version of the dual. 
Computing the rank-based dual can be structured into three parts:
First, we group the nodes and decide in which order these groups should be placed (lines \ref{algo:grouping:1}--\ref{algo:grouping:2}).
Second, we look at each of the groups and sort the nodes by rank, and improve the sorting using \textit{consecutive ones sequences} (\autoref{sec:euler-dual:co}) and distance to the previous set group (lines \ref{algo:CO:1}--\ref{algo:CO:3}).
Finally, we place each node so that it maximizes the number of monotone faces while linking them to the already existing nodes in the graph, removing unwanted crossings (lines \ref{algo:insert:1}--\ref{algo:insert:3}).

\subsection{Grouping by Participating Sets}\label{sec:euler-dual:grouping}

As described in the previous section, each node has one or more participating sets. 
To create the dual, we start by separating these nodes into groups (\autoref{algo:insert:1}).
We first sort the sets by the lowest rank of each set, and the number of nodes they participate in (\autoref{algo:insert:2}). 
In practice, this means that sets that contain nodes with lower ranks such as $\Set{a}$ are considered first.
We then iterate over the individual sets and group nodes that extend the nodes of the previous set with the current set (set extension) in the previously computed order.
An example of this can be seen in \autoref{fig:insertion}a:
Nodes are arranged vertically by rank and the different colors represent the resulting groups from each extension step.
The numbers for each node describe the order in which nodes are added to the groups.
Grouping the nodes using the extension of each set gives us a general order in which nodes are inserted into the dual graph.
For each group, nodes are inserted in a rank-based order. 
Meaning nodes that have a lower rank are placed first. 
However, this on its own is not enough. 
If we were to insert the nodes in the order determined only by their rank and grouping order, the resulting dual will not be planar.
This can be seen in \autoref{fig:insertion}b. 
Here, we are currently inserting the nodes from the red set with rank 2. 
If we naively insert and connect nodes, each node of the rank is connected to all nodes in the rank above using all possible links. 
Therefore, we need to establish the correct order for the nodes within a group, and the correct subset of links, which we will discuss next.

\subsection{Consecutive Ones Sequences}\label{sec:euler-dual:co}

To determine the order of nodes within a group we need to introduce the concept of \textit{consecutive ones} (CO).
Imagine the following scenario: Given an adjacency matrix of a graph, this graph has the consecutive ones property, if we can reorder the rows of the adjacency matrix so that all $1$s in the columns are consecutive.
This property was defined by Booth, and is true for graphs that have a planar embedding~\cite{BoothL76}.
A \textit{consecutive ones sequence} is then a group of consecutive nodes.
As we want to create planar duals, we can use this property in our construction algorithm. 
Remember our goal is to insert nodes into the Euler dual so that as many monotone faces as possible are created.
\st{Here it is important to note that w}We do not need to realize all possible links between nodes.
The only thing that we need to ensure is that the resulting graph is \textit{planar} and \textit{connected}.
In the rank-based dual, it suffices to ensure the CO property for neighboring ranks.
As there are more links in the abstract description than we need for the Euler dual, there are also multiple potential CO sequences, from which we choose the CO sequence that \textbf{maximizes the number of monotone faces}.

If we think back to the overall goal, which is to maximize monotone faces, we can see that the longer the consecutive ones sequence is, the more monotone faces are closed and created, when inserting the corresponding node.
Therefore, we change the order of the nodes on each rank, so that nodes with longer CO sequences are placed before nodes with shorter consecutive ones sequences.
If the length of COs is equal, we further sort the nodes by their respective \textit{twins} in the previous group, without the current set, and sort them by their distance to the closest CO sequence of length 1 of the current group in the rank above (dist\_twin in \autoref{algo:CO:3})
In the example of \autoref{fig:insertion}c, we can see that by reordering the nodes so that we first place node \nodeIcon{redNode}{12} and \nodeIcon{redNode}{9}, and then node \nodeIcon{redNode}{10}, the nodes \nodeIcon{redNode}{12} and \nodeIcon{redNode}{9} will not produce crossings anymore. 

However, as shown in \autoref{fig:insertion}c, some crossings still remain. To adhere to the CO property, all possible CO sequences of a node have to be reduced to a single CO sequence.
For example node \nodeIcon{redNode}{10} has two possible parents nodes in the rank above---nodes \nodeIcon{orangeNode}{2} and \nodeIcon{redNode}{8}. 
The latter two are not adjacent, as they generate two CO sequences.
So, to insert \nodeIcon{redNode}{10}, we have to choose one of the two. 

For each CO sequence and for each possible position in the current rank, we calculate a set of attributes that helps to make the decision where to place it.
These attributes consider the length of the CO sequence, and the change in \#monotone Faces and \#crossings (\autoref{algo:CO:3}).
We collect these attributes across the CO sequences and the possible positions of the node in a list.
As an example, a new node might destroy an existing face, if we place it inside the face. Because the newly inserted node has to be connected to the next rank, a crossing will appear, and the \#monotone faces decreases. 
After we have sorted the list accordingly, we insert the node at the current best position (\autoref{algo:insert:2}), resolve crossings (\autoref{algo:insert:3}), and move on to the next node. 

Returning to our previous example, which can be observed in \autoref{fig:insertion}d, the CO sequence $\Set{d}$ is chosen, because this way no previously created face is destroyed. This is because the node \nodeIcon{redNode}{8} has already been inserted in the rank above, and has created an open space between the nodes \nodeIcon{redNode}{12} and \nodeIcon{redNode}{9}.
We therefore insert the node \nodeIcon{redNode}{10} into this open space, keeping existing monotone faces intact.

Once all nodes of the current rank have been placed, we continue on to the next rank. If we have placed all nodes of the current group, we move on to the next set, get all its nodes, sort the nodes on each rank, and insert the nodes, rank by rank. Using this method, it is possible to create Euler duals that are \textit{connected}, \textit{planar} and contain only \textit{monotone faces}.
We will discuss problematic cases, where this is not possible, in \autoref{sec:discussion}.

\section{Circular Layout of the Euler diagram}\label{sec:circular-layout}

Based on the rank-based Euler dual, we can create the curves of the final Euler diagram.
We do this by first removing the empty set and then arranging the nodes in a circular layout (\autoref{fig:overview:circular-dual}).
At the center of this layout is the intersection with the largest rank, which is usually the full-set.
The other nodes are placed on rings around the center, depending on their rank.
Using this layout, we then devise a strategy to draw smooth curves that result in the final diagram (\autoref{fig:overview:euler-diagram}).

\subsection{Circular Layout}

To guarantee a good distribution of nodes on each ring, we need to place the nodes at well-defined distances to each other.
The rank with the largest number of nodes---usually the middle rank---is placed first to guarantee an overlap-free and well-distributed result.
Rings are then placed outwards and inwards of this rank.
The radii are chosen so that there is still enough space in the inner rings for all nodes, while the outer rings are not too distant from each other. 
Distributing nodes evenly on each ring can result in clutter in the ranks above and below.
Therefore, we place nodes so that enough space is reserved for their children and parents.
Accounting for this, nodes with many children require more space compared to nodes with only a few children.
This approach is similar to the layout of \textit{radial trees}, but with the tree growing in both directions.
The circular layout is then used to create the final Venn diagram.
\autoref{fig:curves:circular-dual} shows the circular distribution of the Euler dual from \autoref{fig:properties:duals:well-matched}.
On each ring, the nodes are well-distributed.

\subsection{Drawing Curves}

To create an Euler diagram from the dual, the simplest approach would be to use the convex hull of the nodes for each set to create a closed shape.
However, such a curve would not consider the nodes outside of the current set. 
This results in closed curves that create many unwanted zones\st{for a single intersection} and a very uneven distribution of areas across the faces.
This is clearly not well-matched and decreases readability.

Therefore, we developed an approach to directly control the curve of each set by introducing additional virtual nodes that act as control points for its shape.
We call these nodes \textit{gate nodes}.
They lie on the same circular path as the intersection nodes but are distributed so that they always lie at the midpoint between two nodes on the ring (\autoref{fig:curves:circular-layout}, dark gray circles).
When we move between ranks, we cross different circular paths in the circular layout, depending on the ranks of the current and following link. 
As we want to control the shape, we define where this crossing is allowed to happen: only at a gate node position.
Due to the properties of the circular graph, which is still a dual of the Euler diagram, we can then create a set curve by finding the order of the links of each set and connecting the midpoints of the links with the gate nodes.
This generates a path that moves between the rings, ``cutting'' the dual graph into two disconnected components.

Using the gate nodes in combination with the midpoints of the links in their respective link order, we create a mostly compact, closed curve for each set.
Additionally, we can control the shape of the curve by using different interpolation strategies and adapting the link-mid point.
We achieved the best curve results using Catmull-Rom splines.
\autoref{fig:curves:circular-layout} shows a circular graph with the shape of a set defined by intersection link midpoints and gate nodes.
Even though we use splines in our approach to maximize the smoothness, it would be possible to constrain the curve further to generate curves that can only use diagonal, rectangular, or octagonal lines. 

\subsection{Concurrency}

\begin{figure}[tb]
    \vspace*{-7mm}
    \centering
    \begin{subfigure}[b]{0.49\linewidth}
        \centering
        \includegraphics[trim=0mm 10mm 0mm 0mm, width = \linewidth]{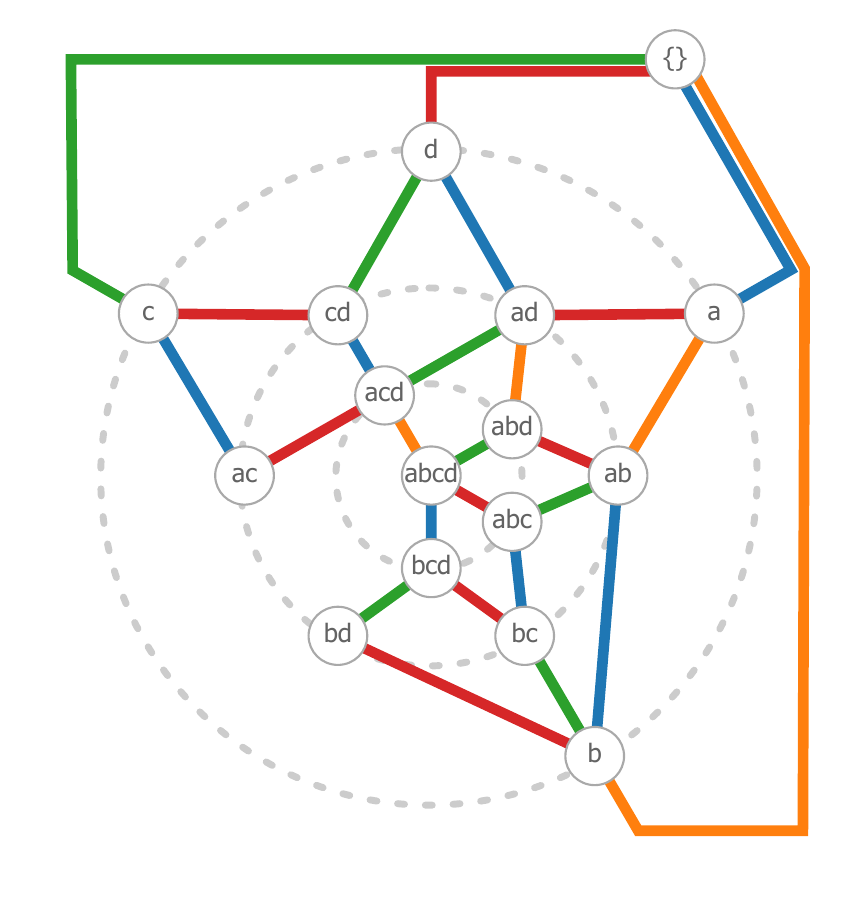}
    \caption{}
    \label{fig:curves:circular-dual}
    \end{subfigure}
    \begin{subfigure}[b]{0.49\linewidth}
        \centering
        \includegraphics[trim=0mm 10mm 0mm 5mm, width =\linewidth]{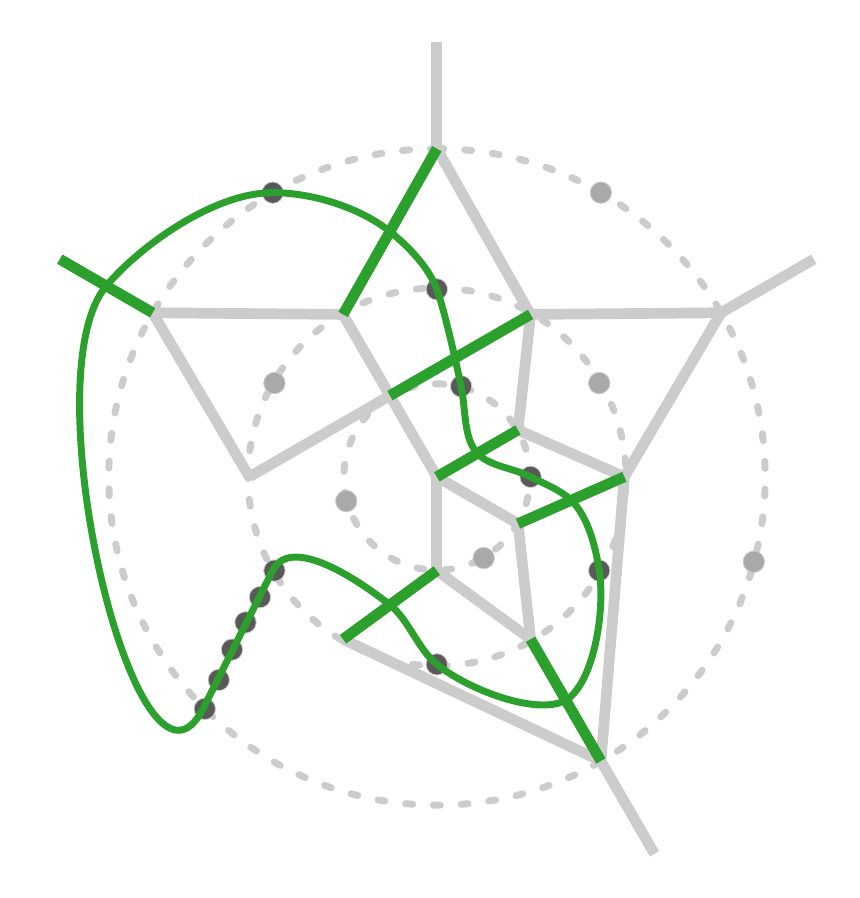} 
    \caption{}
    \label{fig:curves:circular-layout}
    \end{subfigure}
    \caption{To define the shape of the Euler diagram, we order the links for each set along the circles and shape them with gate nodes---shown here as grey dots---between the intersection nodes. This enables us to fine tune their shapes.}
    \label{fig:curves}
\end{figure}

Euler duals that only consist of monotone faces will only have pairwise crossings in the diagram.
However, if we have a non-monotone face, this is not the case.
Instead, we will create non-pairwise crossings.
Curves that use the same gate nodes to cross a ring will produce a concurrent curve segment.
To retain a well-matched diagram, we control the curves, which avoids creating unwanted zones.
This means that for concurrent segments, each curve is offset according to the order in which they enter the concurrent segment.
As the Catmull-Rom interpolation cannot handle straight line segments easily, we instead split the curve into different segments and add additional points to create segments that are concurrent. 
This can be observed in the bottom left part of \autoref{fig:curves:circular-layout}.
If the concurrency is not a straight line but instead happens on the outside of the diagram, we create the curve normally but offset the curves as previously explained.
These are then combined with normal curve segments to create the final closed smooth shape for each set.
\st{We tried different strategies to visualize the concurrency, \eg, alternating stripes and concurrent lines.}

\section{Evaluation}

\begin{figure*}[t]
    \begin{minipage}[h][][b]{0.20\linewidth}
		\begin{subfigure}[b]{0.8\textwidth}
        \includegraphics[valign=t,width=\linewidth]{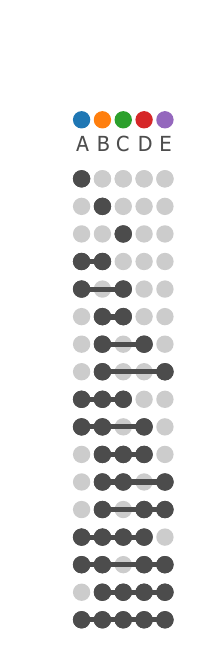}
        \caption{UpSet plot}
    \end{subfigure}
	\end{minipage}
	\begin{minipage}[h]{0.75\linewidth}
	   \begin{subfigure}[b]{0.32\textwidth}
        \includegraphics[width=\linewidth]{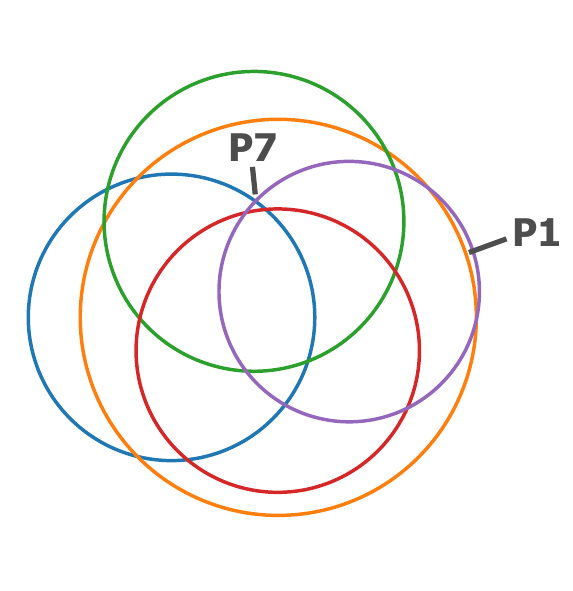}
        \caption{EulerR}
    \end{subfigure}
    \begin{subfigure}[b]{0.32\textwidth}
        \includegraphics[width=\linewidth]{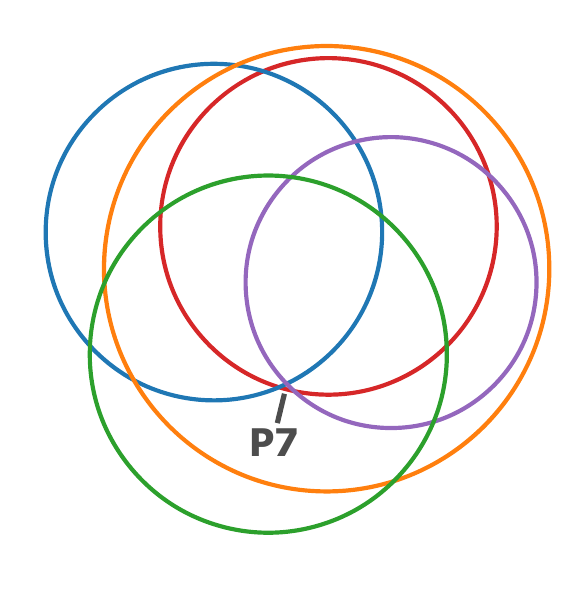}
        \caption{VennEuler}
    \end{subfigure}
    \begin{subfigure}[b]{0.32\textwidth}
        \includegraphics[width=\linewidth]{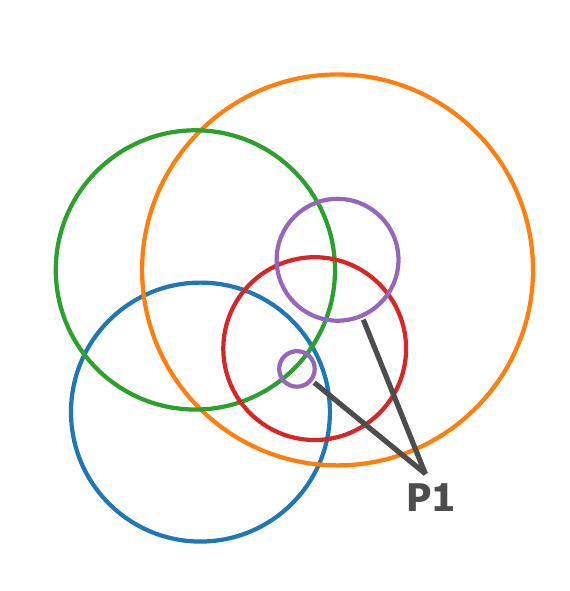}
        \caption{SetNet}
    \end{subfigure}\\
    \begin{subfigure}[b]{0.32\textwidth}
        \includegraphics[width=\linewidth]{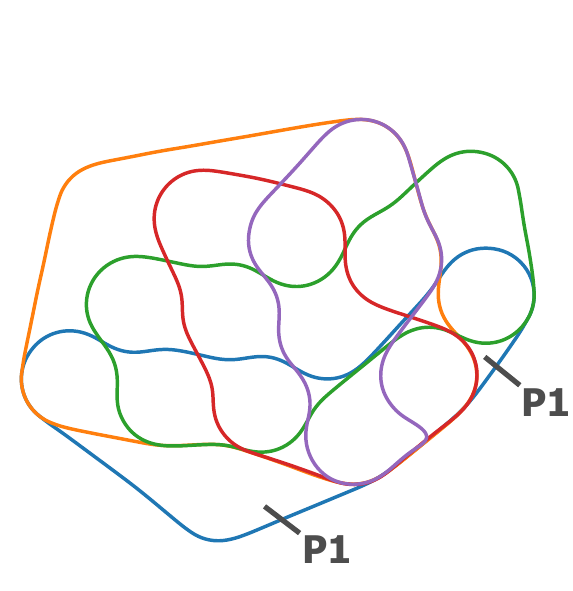}
        \caption{nVenn}
    \end{subfigure}
    \begin{subfigure}[b]{0.32\textwidth}
    \includegraphics[width=\linewidth]{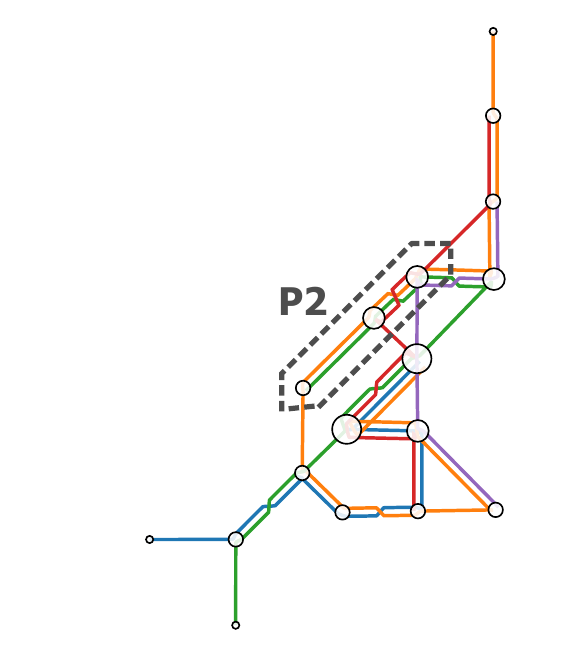}
        
        \caption{MetroSets}\label{fig:topic-modeling:metrosets}
    \end{subfigure}
    \begin{subfigure}[b]{0.32\textwidth}
        \includegraphics[width=\linewidth]{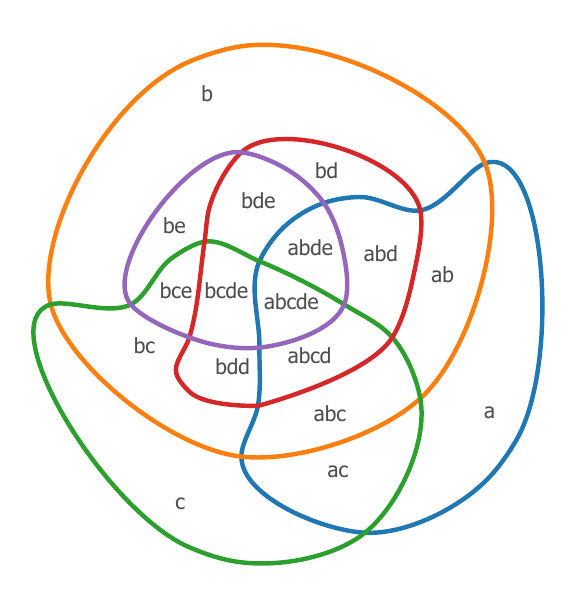}
        \caption{Our method}
    \end{subfigure}
    \end{minipage}
    \caption{Comparing relevant previous works for Euler diagrams of a topic modeling dataset. Problems in the results are marked: these can either be not well-formed (\PropWellFormed{}), not well-matched (\PropWellMatched{}), or create zones that only have very little area (\PropZoneArea{}). Our method produces a result that does not destroy the well-formed and well-matched properties. The areas of the zones are distributed evenly and the shape is compact.
    }
    \label{fig:TopicModelling}
\end{figure*}

\begin{figure*}[t]
    \centering
    \begin{subfigure}[b]{0.2\textwidth}
        \includegraphics[width=\linewidth]{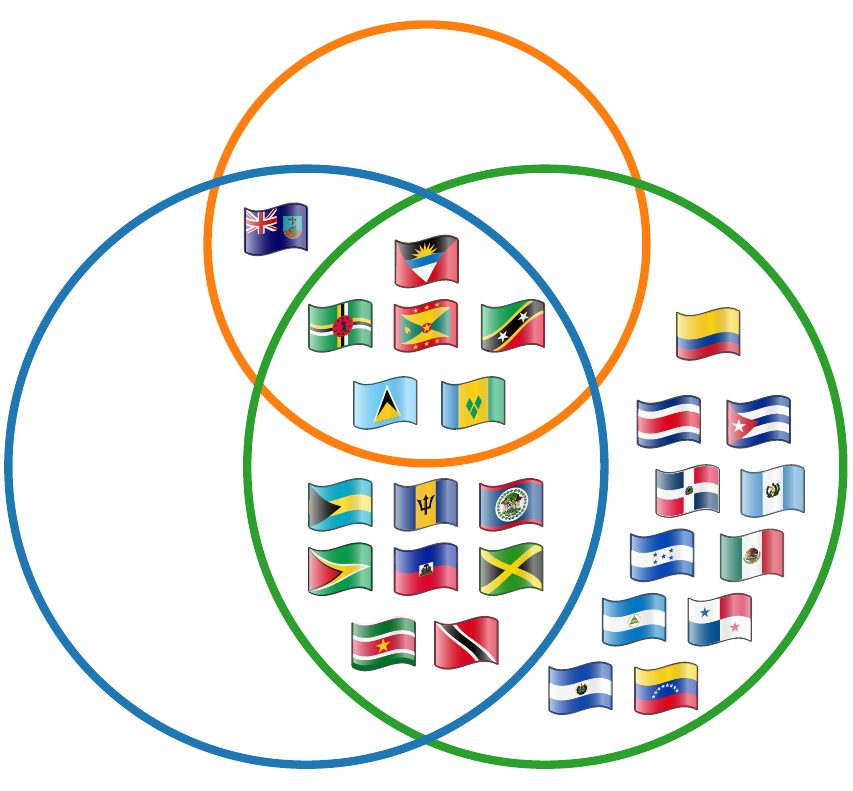}
        \caption{Original}\label{fig:flags:original}
    \end{subfigure}
    \begin{subfigure}[b]{0.2\textwidth}
        \includegraphics[width=\linewidth]{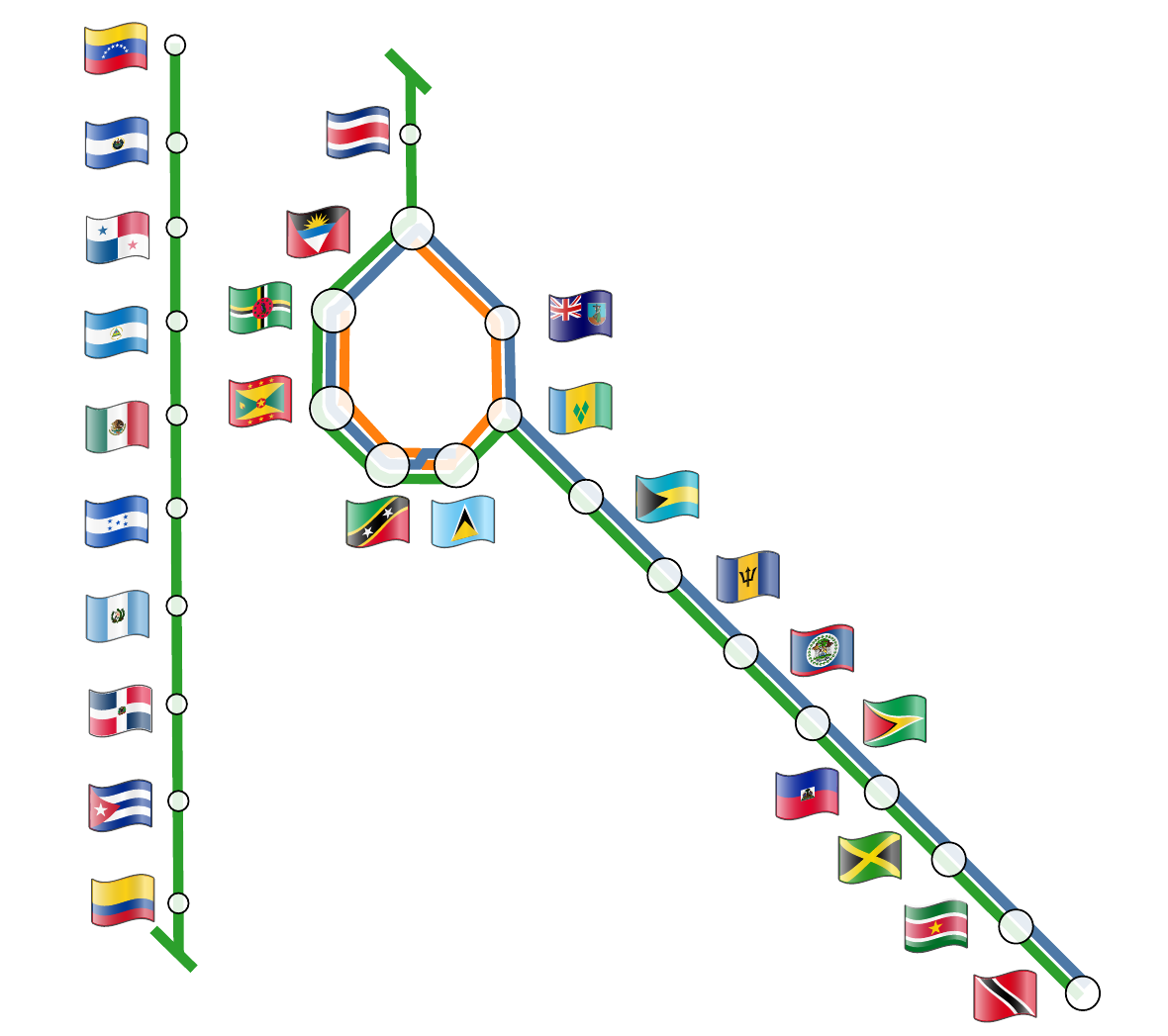}
        \caption{MetroSets}\label{fig:flags:metro-sets}
    \end{subfigure}
    \begin{subfigure}[b]{0.2\textwidth}
        \includegraphics[width=\linewidth]{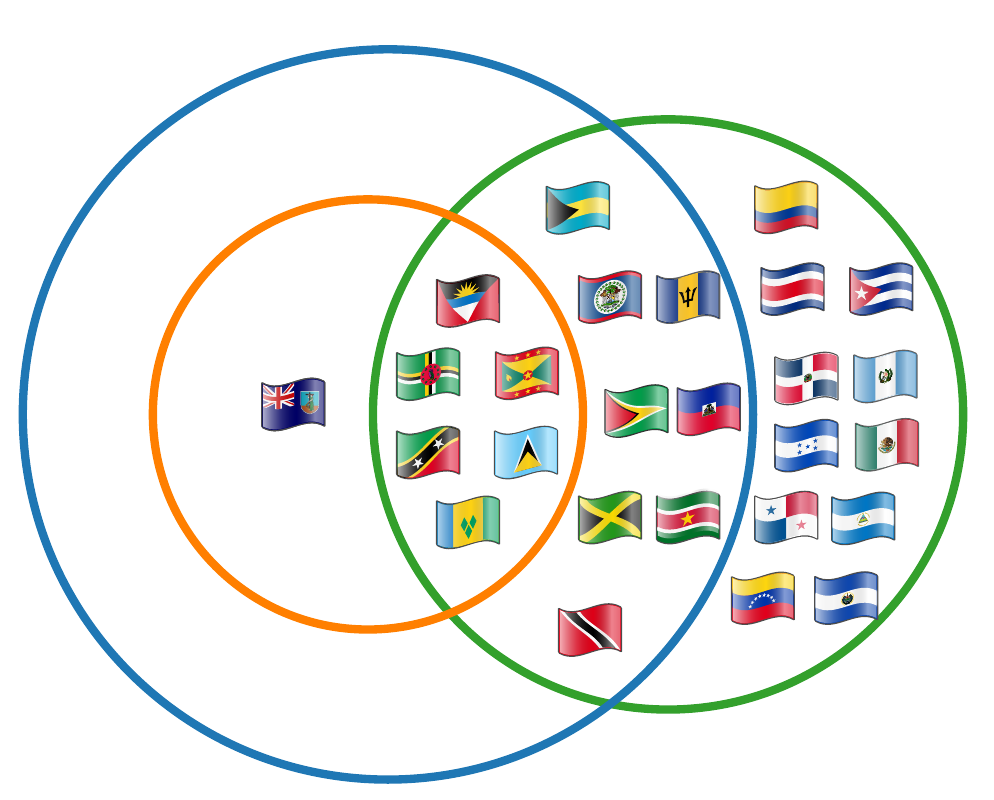}
        \caption{SetNet}\label{fig:flags:setnet}
    \end{subfigure}
        \begin{subfigure}[b]{0.2\textwidth}
        \includegraphics[width=\linewidth]{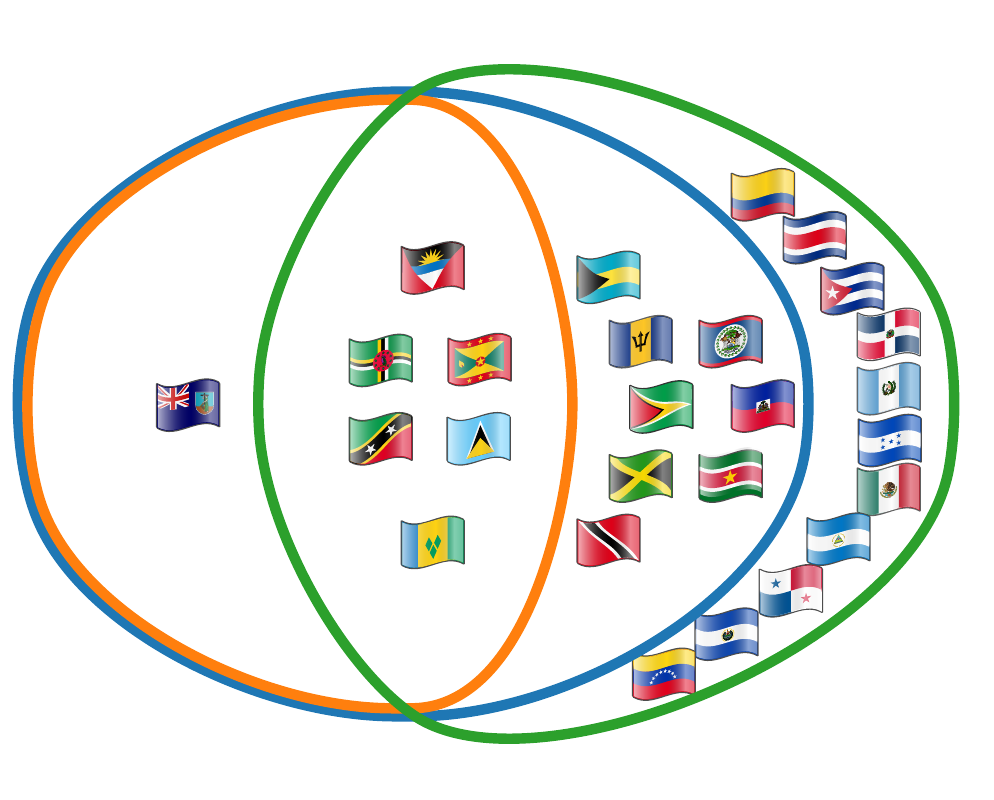}
        \caption{Our method}\label{fig:flags:ours}
    \end{subfigure}
    \caption[]{Here, we show a dataset of infographics concerning different \revised{supranational Caribbean bodies} \st{Caribbean bodies} and their contained countries (a). While MetroSets is well-matched, the visualization requires a lot of space to show all the data points \revised{(b)}\st{(d)}. Therefore, we introduced a discontinuity between \FlagCostaRica{} \textit{Costa Rica} and \FlagColombia{} \textit{Colombia}.
    SetNet introduces unwanted zones, as \Set{a} does not contain any data points (c).
    Our diagram has only a single concurrency and is well-matched (d).
    }\label{fig:flags}
\end{figure*}

We directly compare our method to several other state-of-the-art approaches across three different datasets.
Many older set visualization techniques are not made publicly available~\cite{SimonettoA09,wallinger2021readability}, so it is not possible to compare ourselves directly to them, or they only work on a very limited amount of set curves~\cite{Micallef2014}.
We evaluate our method against vennEuler~\cite{Wilkinson2012}, EulerR~\cite{Larsson2020}, SetNet~\cite{RodgersSAMBT16}, nVenn~\cite{Perez-SilvaAQ18}, and MetroSets~\cite{Jacobsen2021}.
vennEuler, EulerR, and SetNet only allow circles as curves, whereas nVenn allows for arbitrarily shaped curves.
MetroSets are conceptually different from the other techniques as they only produce the Euler graph as an output. \revised{Therefore, we only compared them based on criteria that can be applied to the lines of the graph, such as line intersections, concurrency and overall compactness.} 
Some of these approaches also have an additional weight parameter for each intersection that is used to create area-proportionate diagrams.
Because such factors skew the comparison, we used an equal weight for all set intersections in these methods to unbias the individual areas.
MetroSets allow to directly show data points---examples can be seen in \autoref{fig:flags:metro-sets} and \autoref{fig:topic-modeling:metrosets}\st{, where each circle represents one data point}.
This is not supported by other methods\revised{, including ours}.
To overcome this, the datapoints in \autoref{fig:flags} and \autoref{fig:teaser} were added manually.
As datasets, we chose three examples from different domains: topic modeling and info-graphics. \revised{These datasets show a wide variety in their structure, as well as the kind of datapoints that can be overlayed on top of the diagram.} We will discuss the results according to well-matchedness, well-formedness, as well as additional properties that we are going to introduce in the next section.

\subsection{Guidelines for Euler diagrams}\label{sec:guidelines}

We base our evaluation on the guidelines proposed by Blake et al.~\cite{Blake2016}.
They define 10 different measures which can be used to judge the quality of Euler diagrams, ranked by their importance.
There are three properties that we will not discuss in detail: Diverging lines, orientation, and color. 
Diverging lines are not applicable, and orientation is not considered by any method presented.
It can also easily be changed by rotating the visualization. 
In order to strengthen the comparability of the methods we changed the color and style of all evaluated works to \revised{match ours}\st{be the same}. 
As Blake et al.~\cite{Blake2016} found that only outlines are preferred,\st{ so} we refrain from filling the curves. 
Some of the measures, such as \textbf{well-matchedness} (\PropWellMatched{}) and \textbf{well-formedness} (\PropWellFormed{}), have already been defined in \autoref{sec:properties}.
The others\st{, which we use in our evaluation with other methods,} will be described briefly.
They all relate to the form of the diagram but are not captured in the notion of well-formedness.

\paragraph{Curve Guidelines}
The \textbf{Compactness} (\PropConvex{}) defines how close a shape is to a perfect circle.
Blake et al.~\cite{Blake2016} call this property \textit{shape}, as they only consider circles.
This is closely connected to the convexity of a shape, and there are further studies on the general understanding of convex vs. non-convex shapes~\cite{Hulleman2000, Bertamini2012, Schmidtmann2015}.
In general, they conclude that convexity allows users to finish set-related tasks faster, albeit it might make individual curves harder to distinguish.
\textbf{Smooth curves} (\PropSmoothCurves{}) are preferred by users and result in diagrams that are easier to read.
\st{As a fundamentally different technique, it does not apply to MetroSets.}

\paragraph{Diagram Guidelines}
\textbf{Symmetry} (\PropSymmCurves{}) can also be beneficial if the curves are as symmetric as possible while retaining the features that distinguish individual faces.
This property measures the similarity across all shapes in a uniform way.
Circular approaches will always retain perfect symmetry, while more relaxed shapes might produce symmetric, pseudo-symmetric, or non-symmetric results.
If the diagram is symmetric, finding a given intersection face can be challenging because many faces will have a similar shape. 
Therefore \textbf{shape discrimination} (\PropShapeDisc{}{}) is another important property, which defines the uniqueness of individual faces, and allows for effective search tasks.
\textbf{Zone area equality} (\PropZoneArea{}) measures the area of each face in relation to the other faces.
In general, for Euler diagrams that are not area-preserving, the area of each zone should be as similar as possible. 
Area-preserving Euler diagrams, in contrast, try to adapt the size of faces to be equal to a property, for example, to the number of contained data points (cardinality).
Infringing this property means that users might misinterpret the difference in size as a difference in the cardinality of the face. 

\subsection{Topic Modeling} 

Using \textit{latent Dirichlet allocation}~\cite{blei2003latent}, a common topic modeling algorithm, we extracted 5 topics from a political debate.
The result of such a topic modeling algorithm is usually a list of keywords that describe each topic, together with their probability of belonging to said topic.
We filter keywords to retain words for many combinations of topics while still creating an interesting abstract description
\st{\footnote{We will publish the abstract description for this dataset along with our tool.} }
that has a well-matched and well-formed diagram.

One common problem of topic modeling results is that it is very hard to visually compare them just using their descriptive keywords.
Often words are attributed to multiple topics, but just representing them as a list, one cannot easily discern this.
These words, however, might be of special interest to the user.
They might describe all the topics very well, in which case the topics might be very similar to each other, or they might be general "common" words that should not be considered by the topic model algorithm, as they reduce the descriptiveness.
In this section, we only show the resulting curves; the full diagram including the words can be found in the supplemental material.

\autoref{fig:TopicModelling} shows the results for the above dataset across all methods.
For easier comparison, we have highlighted problematic zones in the respective diagrams, which result from infringements of the well-matchedness (\PropWellMatched{}), well-formedness (\PropWellFormed{}), and area-equality (\PropZoneArea{}) properties.
Only a subset of the infringements is shown, as the diagrams might otherwise become unreadable.
Some approaches are very similar (vennEuler and EulerR), while others diverge substantially.
\st{We will now give a small summary of each method and identify potential problems.}

Most methods preserve the abstract description faithfully (\PropWellMatched{}).
However, both EulerR and nVenn create intersections that do not appear in the abstract description.
nVenn even realizes some relations with multiple faces, that appear in different parts of the visualization.
Regarding well-formedness (\PropWellFormed{}), we can observe that all circular visualizations (b-d) are simple.
However, this comes at a cost: SetNet creates \revised{duplicate}\st{duplicates} curves for $\Set{E}$, while the other two approaches are not well-matched.
nVenn and MetrosSets are not simple, as they contain non-pairwise crossings and concurrency.
vennEuler, SetNet, and EulerR use circles and are therefore perfectly compact (\PropConvex{}).
But nVenn also produces relatively compact shapes.
MetroSets, on the other hand, produce a very spread out intersection graph that does not fit into a compact shape.
All results that produce Euler diagrams create smooth curves (\PropSmoothCurves{}).
Symmetry is not considered in any of the related work (\PropSymmCurves{}).
Circular methods create zones that are easily distinguishable, whereas the zones produced by nVenn are very similar to each other.
As MetroSets only create intersection nodes, no real shape is created that can be considered here (\PropShapeDisc{}). All of EulerR, VennEuler, SetNet, and nVenn create very small zones that are difficult to recognize (\PropZoneArea{}).

Our method produces a both well-matched (\PropWellMatched{}) and well-formed (\PropWellFormed{}) result.
The resulting shapes are mostly compact (\PropConvex{}).
As we use curve interpolation, the \revised{produced}\st{produces} curves are smooth (\PropSmoothCurves{}).
In our method, some curves retain their symmetry at least partly (\PropSymmCurves{}).
However, because of this, the zones are also more similar, affecting how easy they are to distinguish (\PropShapeDisc{}).
The area of the zones remains relatively equal across all ranks (\PropZoneArea{}).
In summary, our method retains the guidelines better than all other related works, except for zone discrimination, where we lie between nVenn and vennEuler/EulerR.
Most important, the result is a well-matched and well-formed diagram.

\subsection{Size Venn Diagram}

As a second example, we show a Venn diagram published on \texttt{xkcd.com}\footnote{By \textsc{Randall Munroe} at \texttt{\url{https://xkcd.com/2122}} \ccbync} in \autoref{fig:teaser:original}, that describes possible combinations of words in combination with five different adjectives: \textit{little}, \textit{large}, \textit{small}, \textit{great}, and \textit{big}.
The original visualization uses a 5-Venn diagram to show which words can occur together with these adjectives.
However, there are some combinations for which no words were specified, such as \textit{little}, \textit{large} and \textit{great}.

We can recreate the symmetric 5-Venn diagram used by the author using our algorithm, as can be seen in \autoref{fig:teaser:our-venn}.
The words for each relation are added manually on top of the generated layout.
This gives us the direct equivalent to the hand-made Euler diagram by the author.
Then, we can remove the empty intersections and instead create a well-matched (\PropWellMatched{}) Euler diagram.
In this case, the result is not well-formed (\PropWellFormed{}), so we retain minimal concurrency as well as one non-pairwise intersection.
The resulting curves are mostly compact (\PropConvex{}) and smooth (\PropSmoothCurves{}).
As only a few relations are empty, the diagram retains its high symmetry (\PropSymmCurves{}), but in turn, many zones are similarly shaped (\PropShapeDisc{}).
The area is evenly distributed across the zones (\PropZoneArea{}).
A comparison across the related works can be found in the supplemental material.

\subsection{Supranational Caribbean Bodies}

We recreate another info-graphic visualization published on Wikipedia commons\footnote{By \textsc{Wdcf} at \texttt{\url{https://w.wiki/39HJ}} \ccbysa}, where countries are grouped by organizations.
In this case, we look at all Caribbean countries that are contained in Supranational Caribbean Bodies.
There are three different bodies: the \textit{Association of Caribbean States}, the \textit{Caribbean Community}, and the \textit{Organization of Eastern Caribbean States}.
However, not all intersections between the three exist, as there are no countries for some relations.
The original visualization uses a 3-Venn diagram to visualize the relations.
Existing relations are filled with flags that represent each country.
This makes the visualization quite large, as a lot of space is needed to visualize the empty intersections, even though no data is shown.

As an additional comparison, we \st{also }show how SetNet and MetroSets visualize this data set.
In \autoref{fig:flags:setnet}, we can observe that SetNet does not always preserve well-matchedness (\PropWellMatched{}), as an empty zone is created.
SetNet handles this by placing red dots inside \st{the }faces that are part of the abstract description.
Since we already show the flags of the countries that belong to each intersection directly\st{ in the area}, we chose to omit this in our recreation. 
MetroSets (\autoref{fig:flags:metro-sets}) shows all the data points, in this case countries, directly in the visualization.
However, the visualization needs a lot of space, as lines extend outwards, resulting in a non-compact shape (\PropConvex{}). 

Using our technique, we can visualize the relations as a well-matched (\PropWellMatched{}) Euler diagram.
The diagram has concurrency, as some bodies do not contain countries that are only in this body, and is therefore not well-formed (\PropWellFormed{}).
Curves are compact (\PropConvex{}) and smooth (\PropSmoothCurves{}).
The symmetry is limited (\PropSymmCurves{}), but still the zones are similar (\PropShapeDisc{}).
The area is evenly distributed across the zones (\PropZoneArea{}).
Our visualization allows the reader to immediately see that there are four relations in total.
The central intersection is shared for all three bodies, while there is a single relation that has outer concurrency.

\section{Discussion and Future Work}\label{sec:discussion}

First, we will discuss runtime, problematic abstract descriptions, and alternative construction methods.
Then we will further analyze the influence of design decisions on the aesthetics of the visualization.

\begin{figure}[tb]
    \centering
      \begin{subfigure}[b]{0.35\linewidth}
        \centering
        \includegraphics[width =0.98\linewidth]{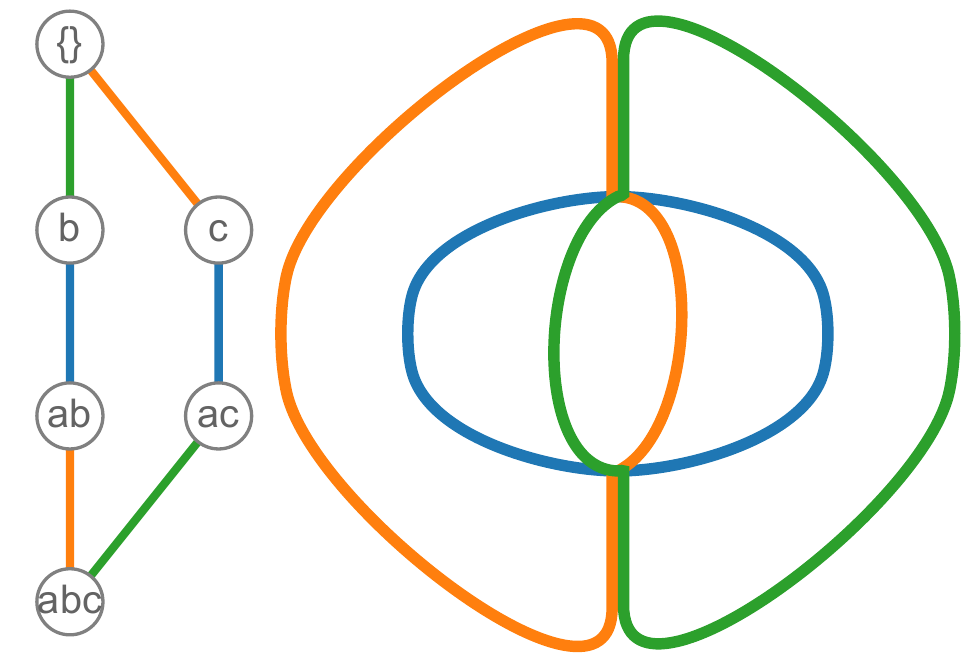}
    \caption{non-monotone faces}
    \label{fig:non-monotone:no-quad}
    \end{subfigure}
    \begin{subfigure}[b]{0.43\linewidth}
        \centering
        \includegraphics[width =0.98\linewidth]{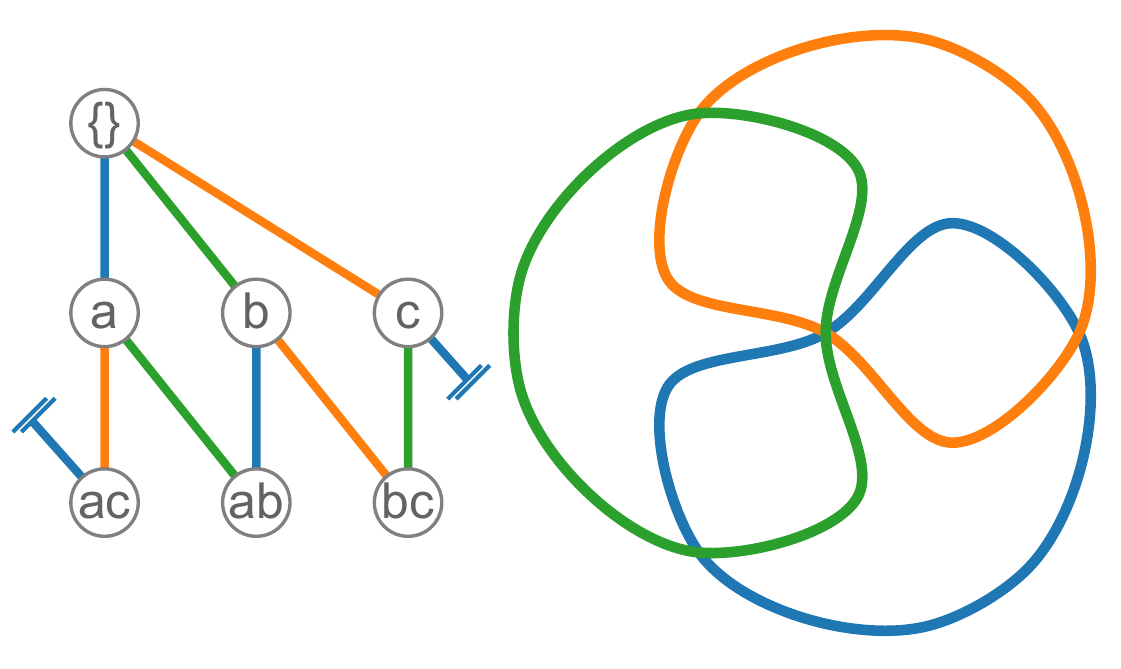}
    \caption{no common sink}
    \label{fig:non-monotone:no-sink}
    \end{subfigure}
    \caption{Examples with problematic abstract descriptions: (a) non-monotone faces will result in complex Euler diagrams. (b) If there is no shared intersection on the highest rank\st{ (sink)}, a non-pairwise intersection will appear in the center of the diagram.} 
    \label{fig:non-monotone}
\end{figure}

\subsection{Runtime}


We performed experiments to compare the runtime of our approach to two other state-of-the-art methods.
Because of its optimization strategy, MetroSets does not scale well with the number of nodes \st{but instead}\revised{and} increases in quadratic time~\cite{Jacobsen2021}.
SetNet extends the \textit{iCircles}~\revised{\cite{StapletonFRH12}} algorithm and runs in polynomial time. 
For lower node counts, our \revised{algorithm}\st{algorithms} has similar runtime ($n=64$, $0.072$s) to SetNet ($n=64$, $0.14$s), whereas MetroSets is a bit slower ($n=64$, $~2$s)\footnote{All experiments were run on a desktop PC with an Intel i5-8400 CPU.}.
\revised{For larger number of intersection nodes, we can still achieve fast results ($n=1024$, $27.48$s).
Our approach is a greedy algorithm that uses grouping and reordering to reduce the search space of possible positions for a new node.
This allows us to avoid complex optimization strategies, making the output deterministic.}
\st{The reason for this is that by using our reordering strategy and restricting the search space of possible positions for a new node, we do not need to solve
complex optimization tasks. As our method only needs to do an initial sorting of all nodes and afterwards is constrained to nodes of the same rank, we can achieve very fast results ($n=1024$, $0.72$s).}
From our experiments, we expect our algorithm to run in polynomial time 
with the grouping of nodes (\autoref{sec:euler-dual:grouping}) as the limiting factor.
However, we hope to prove stronger bounds for this in future work.


\subsection{Problematic Abstract Descriptions}\label{sec:problematic-descriptions}

As we have discussed before in \autoref{sec:properties}, the layout of an Euler diagram strongly depends on the abstract description, and in particular, if there exists a well-formed solution for it.
Our method handles this by relaxing the well-formedness properties if otherwise no such diagram can be found while guaranteeing the well-matchedness property.
In contrast, other related works, such as eulerR, vennEuler, nVenn, or sometimes even SetNet, fail silently for these abstract descriptions or, arguably worse, create diagrams that are not well-matched.
As we rate well-matchedness above all other attributes, this usually means for problematic abstract descriptions that we create non-monotone faces.
An example of this is shown in \autoref{fig:non-monotone}.\st{, where non-monotone faces result in concurrent curves and non-pairwise intersections in the Euler diagram (\autoref{fig:non-monotone:no-quad}). }
If there is no common intersection for all sets, all sets will intersect in the center of the diagram and cause a non-pairwise intersection(\autoref{fig:non-monotone:no-sink}).
If there is no monotone connection between the empty set and the lowest-ranked nodes, the outer curve will have concurrent curves, as can be seen in \autoref{fig:flags:ours}.

There is one aspect of abstract description that we have not considered so far:
In some cases, it can happen that the resulting diagram as a whole could be disconnected, or only some sets will have a disconnect, meaning nodes cannot be connected via a strong monotone link to at least one parent node and one child node.
\revised{These datasets can currently not be visualized in our tool.}
We plan to remove this limitation in future work, by having special solutions for these disconnected components, for example, separation of the nodes so that each disconnected component is visualized by its own curve or concurrency on the rings, similar to the method of collapsing faces \st{proposed} by Chow~\cite{ChowR05}.

Finally, we want to point out that although our algorithm always produces a valid well-matched result, so far we have not proven that this result is optimal with respect to minimal violations of the well-formedness property.
We also aim to pursue this in future work.


\subsection{Alternative Construction Methods}\label{sec:alternative-construction}

Initially, we also tried alternative construction methods for Euler diagrams.
We adapted the backtracking algorithm that Mamakani et al.~\cite{MamakaniMR12} proposed for finding symmetric Venn diagrams to handle arbitrary Euler diagrams.
The basic idea behind this technique is to find suitable crossings by permuting the order of curves.
We used this approach to investigate the proportion of abstract descriptions for which well-formed diagrams exist.
The number of possible intersections grows drastically with each additional set:
For $n = 4$ there are 65K+ combinations---for $n = 5$ there are already 4M+ possible combinations.
Furthermore, we only consider descriptions where each node has at least one incoming and one outgoing link, all sets have a common source as well as sink, and the empty set always exists. 
As described in \autoref{sec:problematic-descriptions}, these are the abstract descriptions for which the diagram is connected and monotone.
This lowers the combinations significantly to 3152 for $n=4$.
Using the modified backtracking, we found that for $n=4$, only 125 out of the 3152 combinations have monotone diagrams (3,96 \%).
In the rest of the cases, the backtracking approach would find no solution at all.
The reason for this is simple:
Backtracking cannot readily find sub-optimal solutions, which motivates our proposed method.
By allowing non-monotone faces as the last resort, our approach always yields a valid Euler diagram.


\subsection{Design Considerations}

\paragraph{Curves}
When we create the diagram using our method, we use customized splines to create a single, smooth curve for each set.
We have also performed experiments using convex hulls and linear poly-hulls for drawing curves, but these approaches cannot guarantee well-formedness, which is why we chose Catmull-Rom curves.
In particular, we segment the curve into parts and use different strategies depending on what kind of curve segment occurs.
There are three different types of segments: regular segments, U-turn segments, and concurrent segments.
This differentiation allows us to be flexible in our choice of interpolation strategies, and we can control the smoothness of the curve by adapting the number of control points.
For instances where concurrency cannot be avoided, we tried different approaches to mitigate it, \eg, \st{true} concurrent\st{ly running} lines or thinner lines so that equal line width is retained. 
Another approach would be \st{to create} dashed stroke segments with alternating colors.
\st{We plan to allow for more constrained curve shapes, such as octolinear or rectangular shapes, in the future.}

Another interesting consideration is how to visualize the individual intersections.
\st{Usually, w}\revised{W}e \st{only} show the curves without\st{ any} filled-in areas~\cite{Blake2016}.
The problem with filling the areas of the curves is that, because of blending, each intersection will have a unique color that is not part of the original color set.
With \st{an }increasing number of sets, the visual difference between these colors decreases, which makes it hard to distinguish the intersections.
Methods have been developed to alleviate these effects~\cite{alsallakh2014visual}, but we consider their application \st{and evaluation }out of the scope for this paper.

\revised{A limitation of using Catmull-Rom curves is that for large abstract descriptions\st{with many set intersection nodes}, curves might become complex and non-convex, which might have a negative impact on the readability of the diagram. This effect can already be observed Fig.~\ref{fig:teaser:our-euler}, and it increases for highly-intersecting datasets, which can also be seen in the supplementary.
Solutions to this problem could either be to directly adapt the Euler dual by reordering nodes, or to post-process the faces of the diagram by optimizing the convexity of their outline. }

\paragraph{Area-proportionate Diagrams}
\revised{Cardinality is an important characteristic that is inherent to the data, if it exists in a given dataset. 
Currently, our method does not incorporate information about the cardinality of set intersection nodes.}
In the future, we \st{also} plan to integrate a method to adapt the zones to given data point weights, creating area-proportionate diagrams, and fill these zones automatically with data points. \revised{This will also remove a current limitation of our tool, as it does not scale with large number of datapoints in a single zone.}

\paragraph{Symmetry}
Another factor that influences the aesthetics of the diagram is symmetry, which many approaches do not retain.
However, we believe that symmetry is an important aspect to investigate with regard to user engagement.
Symmetric objects are often perceived as more aesthetic~\cite{Cawthon2007}, especially so rotational symmetry~\cite{Makin2012}.
However, striving for symmetry goes against the guidelines by Blake et al.~\cite{Blake2016}, which we introduced in \autoref{sec:guidelines}, as symmetry leads to zones that are visually very similar to another.
\st{Similar shapes}\revised{These} violate the shape discrimination property (\PropShapeDisc{}) and can hinder user-understanding.
An extreme example of this is nVenn.
Ultimately, we believe that this is a trade-off that has to be made depending on the goal of the visualization.
For efficiency, the diagram should be simple and easy to read.
If, on the other hand, the goal is to achieve an aesthetic representation, a more symmetric result can be chosen, 
\revised{based on user preference.}
By default, our method creates well-matched diagrams.
Depending on the task, it can also make sense to draw empty intersections to emphasize the absence of instances.
An example of this can be seen in \autoref{fig:teaser:our-venn}.

Using our method, it is also possible to create well-matched and well-formed Venn diagrams for any number of curves very fast. This is because of the inductive nature of Venn diagrams. Because each new set in a Venn diagram doubles the number of nodes, we can simply for each rank inverse the order of the nodes in the rank before, extend the nodes using the new set, and insert the nodes and links.
However, currently some artifacts appear at higher set counts ($n\geq8$). This is because the space of the individual faces is compressed a lot. We plan to improve our curve interpolation to handle Venn diagrams with more curves. The current problems can be observed in the supplementary material. 
A more difficult problem is creating symmetric Venn diagrams. Currently, we are able to create symmetric Venn diagrams for 5 and 7 sets by constricting the number of children a node can have. This influences the insertion strategy so that CO sequences are preferred that have fewer children while still maximizing monotone faces. 
\st{Currently, i}It is still an open problem to find simple symmetric Venn diagrams for any prime number of sets, and the largest to be produced is a 13-Venn diagram \cite{Mamakani2014}.
\section{Conclusion}

We have presented \textsc{\SystemName}\st{, semantics-preserving Euler diagrams}, a novel approach to create well-matched Euler diagrams that focuses on creating mostly simple, planar, and connected solutions.
The visualization of highly connected sets is a challenging problem, as the number of possible intersections increases exponentially with the number of sets.
Our\st{ deterministic} solution is fast and\st{ creates layouts without any time-consuming optimization strategies.
Its flexibility allows to create Venn and Euler diagrams, while scaling to a large number of sets.}
\revised{ can scale to a large number of intersection nodes.}
We \st{can }imagine \st{that }our technique and accompanying visualization to \st{also }be used as a part of larger \revised{a system}\st{systems} that gives a brief and intuitive overview of set-typed data.

\acknowledgments{
This work was supported by the German Research Foundation (DFG) within project KE 740/17-2 of the FOR2111 “Questions at the Interfaces” as well as Project-ID 251654672 – TRR 161 "Quantitative methods for visual computing". We would further like to thank Matthias Albrecht for help during the implementation and Patrick Paetzold for valuable feedback during the revision process.}

\bibliographystyle{abbrv-doi-hyperref-narrow}

\bibliography{textvis}
\end{document}